\documentstyle[twocolumn,prc,aps,epsfig]{revtex}

\input epsf
\newcommand{\be}{\begin{eqnarray}}
\newcommand{\ee}{\end{eqnarray}}

 \newcommand{\gsim}{\mathrel{\hbox{\rlap{\lower.55ex \hbox {$\sim$}}
                   \kern-.3em \raise.4ex \hbox{$>$}}}}
\newcommand{\lsim}{\mathrel{\hbox{\rlap{\lower.55ex \hbox {$\sim$}}
                   \kern-.3em \raise.4ex \hbox{$<$}}}}

\def\N{${\cal N}\,\,$}

\def\roughly#1{\mathrel{\raise.3ex\hbox{$#1$\kern-.75em%
\lower1ex\hbox{$\sim$}}}}
\def\lsim{\roughly<}
\def\gsim{\roughly>}

\begin{document}

\twocolumn[\hsize\textwidth\columnwidth\hsize\csname @twocolumnfalse\endcsname
\title{Strongly coupled plasma with electric and magnetic charges}
\author {Jinfeng Liao and Edward Shuryak}
\address {Department of Physics and Astronomy, State University of New York,
Stony Brook, NY 11794}
\date{\today}
\maketitle
\begin{abstract}
A number of theoretical and lattice results lead us to believe
that Quark-Gluon Plasma not too far from $T_c$ contains not only
electrically charged quasiparticles -- quarks and gluons -- but
magnetically charged ones -- monopoles and dyons -- as well.
Although binary systems like charge-monopole and charge-dyon were
considered in details before in both classical and quantum
settings, it is the first study of coexisting electric and
magnetic particles in many-body context. We perform Molecular
Dynamics study of strongly coupled plasmas with $\sim 1000$
particles and different fraction of magnetic charges. Correlation
functions and Kubo formulae lead to such transport properties as
diffusion constant, shear viscosity and electric conductivity: we
compare the first two with empirical data from RHIC experiments as
well as results from AdS/CFT correspondence. We also study a
number of collective excitations in these systems.
\end{abstract}
\vspace{0.1in}
]
\newpage

\section{Introduction}

This paper has two main goals. One is to introduce a new view of
finite $T-\mu$ QCD based on a competition between $electrically$
and $magnetically$ charged quasiparticles (to be referred to as
EQPs and MQPs below). The second is to begin quantitative studies
of many-body effects based on these ideas, starting with the
simplest approach possible, namely by use of classical mechanics
and basic forces acting between them.

The motivations and some details related to those quasiparticles
will be explained in the rest of the Introduction: but before
doing so let us start with a short summary of the picture
proposed. It is different from the traditional approach, which
puts $confinement$ phenomenon at the center of the discussion,
dividing the temperature regimes into two basic phases: (i)
confined or hadronic phase at $T<T_c$, and (ii) deconfined or
quark-gluon plasma (QGP) phase at $T>T_c$.

We, on the other hand, focus on the competition of EQPs and MQPs
and divide the phase diagram differently, into (i) the
``magnetically dominated'' region at $T<T_{E=M}$ and (ii)
``electrically dominated'' one at $T>T_{E=M}$. In our opinion, the
key aspect of the physics involved is the {\em coupling strength}
of both interactions. So, a divider is some {\em E-M equilibrium}
region at intermediate $T$-$\mu$. Since it does $not$ correspond
to a singular line, one can define it in various
ways\footnote{Another
  possibility discussed in section \ref{sec_boundstates}
is the curve of marginal stability for a gluon. }: the most direct
one is to use a condition that electric (e) and magnetic (g)
couplings are equal
\be e^2/{\hbar c}= g^2/{\hbar c}= 1/2 \ee
The last equality follows from the
 celebrated Dirac quantization condition \cite{Dirac}
\begin{equation} \label{Dirac_quantization}
\frac{eg}{\hbar c} = \frac{n}{2}
\end{equation}
with $n$ being an integer, put to 1 from now on.

Besides equal couplings, the equilibrium region is also
presumably characterized by comparable densities as well as masses
of both electric and magnetic quasiparticles\footnote{Let us
however remind the reader that the E-M duality is of course not
exact, in particular EQPs are gluons and quarks with spin 1 and
1/2 while MQPs are  spherically symmetric ``hedgehogs'' without
any spin.}.

The ``magnetic-dominated'' low-$T$ (and low-$\mu$) region (i) can
in turn be subdivided into the $confining$ part (i-a) in which
electric field is confined into quantized flux tubes surrounded by
the condensate of MQPs, forming t'Hooft-Mandelstamm ``dual
superconductor''
 \cite{t'Hooft-Mandelstamm}, and a new {\em ``postconfinement''}
 region (i-b) at $T_c<T<T_{E=M}$ in which EQPs are
still strongly coupled (correlated) and still connected by the
electric flux tubes. We believe this picture better corresponds to
a situation in which string-related physics is by no means
terminated at $T=T_c$: rather it is at its maximum there. Then if
leaving this ``magnetic-dominated'' region and passing through the
equilibrium region by increase of $T$ and/or $\mu$, we enter
either the high-$T$ "electric-dominated" QGP or a color-electric
superconductor at high-$\mu$ replacing the dual superconductor
(with diquark condensate taking place of monopole condensate).

\begin{figure}[t]
  \epsfig{file=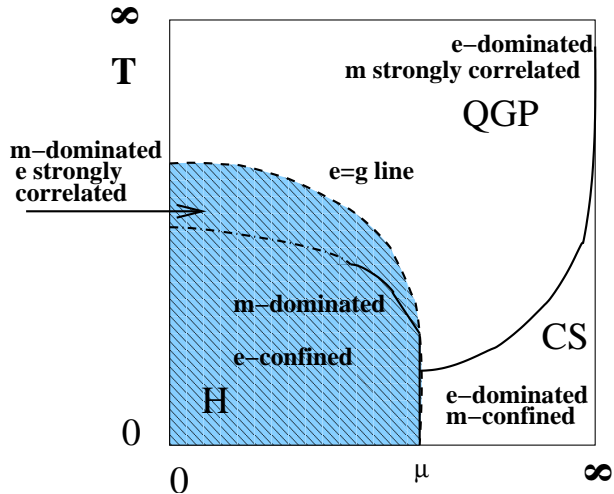,width=8.cm}
  \vspace{0.1in}
\caption{(color online) A schematic phase diagram on
 a (``compactified'') plane of
 temperature and baryonic chemical potential $T-\mu$.
 The (blue) shaded
region shows ``magnetically dominated'' region $g<e$, which
includes the e-confined hadronic phase as well as ``postconfined''
part of the QGP domain. Light region includes ``electrically
dominated'' part of QGP and also color superconductivity (CS)
region, which has e-charged diquark condensates and therefore
obviously m-confined. The dashed line called ``e=g \,line'' is the
line of electric-magnetic equilibrium.
 The solid lines indicate true phase transitions,
while the dash-dotted line is a deconfinement cross-over line.\\ }
\label{fig_em_phasediag}
\end{figure}

A phase diagram explaining this viewpoint pictorially is shown in
Fig.\ref{fig_em_phasediag}.

The paper is structured as follows. We start with a mini-review of
the subject, including RHIC phenomenology and lattice data
relevant for this picture. Then we start discussing classical
dynamics of electric and magnetic charges interacting with each
other. We briefly summarize what is known about the two-body
problems in section II, and proceed to (idealized) 3-body problem,
namely a motion of a magnetically charged object in a field of a
static electric dipole, and discuss flux tubes in classical
plasmas. We then proceed to Molecular Dynamics (MD) simulations.
The main parameters of the model are (i) the ratio of MQPs/EQPs
concentration and (ii) the ratio of magnetic-to-electric coupling
$g/e$. The main issues discussed are how the transport properties
(in particular the $shear$ $viscosity$) of the plasma depend on
them. More specifically, the issue is whether admixture of
weaker-coupled MQPs increases or decreases it.\\

\subsection{Strongly coupled Quark-Gluon plasma in
heavy ion collisions}
\label{sec_boundstates}
A realization~\cite{Shu_liquid,SZ_rethinking} that
QGP at RHIC  is not a weakly coupled
gas but rather a {\em strongly coupled liquid}
 has led to a  paradigm
 shift in the field. It was extensively debated  at
the ``discovery''  BNL workshop in 2004~\cite{discovery_workshop} (at which the
abbreviation sQGP was established) and multiple other meetings since.

Collective flows, related with explosive behavior of hot matter,
were observed at RHIC and studied in detail: the conclusion is
that they are reproduced by the ideal hydrodynamics remarkably
well. Indeed, although these flows affect different secondaries
differently, yet their spectra are in quantitative agreement with
the data for all of them, from $\pi$ to $\Omega^-$. At non-zero
impact parameter the original excited system is deformed in the
transverse plane, creating the so called elliptic flow
 described by  \be v_2(s,p_{\,t},M_i,y,b,A) =<cos(2\phi)> \ee
where $\phi$ is the azimuthal angle and the others stand for the
collision energy, transverse momentum, particle mass, rapidity,
centrality and system size. Hydrodynamics explains  all of those
dependence, for about 99\% of the particles\footnote{The remaining
$\sim 1 \%$ resigning at larger transverse momenta $p_t>2 GeV$ are
influenced by hard processes and jets.}.

Naturally, theorists want to understand the nature of this
behavior by looking at other fields of physics which have prior
experiences with liquid-like plasmas. One of them is related with
the so called AdS/CFT correspondence between strongly coupled
$\cal N$=4 supersymmetric Yang-Mills theory (a relative of QCD) to
weakly coupled string theory in Anti-de-Sitter space (AdS) in
classical SUGRA regime. We will not discuss it in this work: for a
recent brief summary of the results and references see e.g.
\cite{Shu_adv06}.

Zahed and one of us~\cite{SZ_rethinking} argued that marginally bound
states create resonances which can strongly enhance transport
 cross section. Similar phenomenon does happen for ultracold trapped atoms,
 due to Feshbach-type resonances at which the binary scattering length
$a\rightarrow \infty$, which was indeed shown to lead to a
near-perfect liquid. van Hees, Greco and Rapp\cite{Rapp_vanHees}
studied $\bar q c$ resonances, and found enhancement of charm
stopping.

Combining lattice data on quasiparticle masses and interparticle
potentials, one finds a lot of quark and gluon bound states
\cite{SZ_bound,Liao_1} which contribute to thermodynamical
quantities and help explain the ``pressure puzzle''
\cite{SZ_bound}, an apparent contradiction between heavy
quasiparticles near $T_c$ and rather large pressure.
 The magnetic sector discussed in this paper provides
another contribution, that of MQPs (monopoles and dyons),
which will help to resolve the pressure puzzle.

A very interesting issue is related with counting\footnote{And
prevention of the double counting.} of
 the bound states of all quasiparticles. Here the central
notion is that of
{\em curves of marginal stability} (CMS), which are not
 thermodynamic
 singularities but lines indicating a significant change of physics
 where a switch from one language to another (like E$\rightleftharpoons$M )
 is appropriate or even mandatory.

Let us mention one example related with quite interesting
  ``metamorphosis'' discussed in literature,
 in the context of $\cal N$=2 SUSY theories. The CMS
in question is related with the following reaction
\be gluon \leftrightarrow monopole+dyon \ee
in which the r.h.s. system is magnetically bound pair (obviously with
  zero total magnetic charge). The curve itself
is defined by the equality of thresholds, \be M(gluon)=M(dyon)+M(monopole) \ee
As discussed in details by Ritz et al \cite{RV}  ,
inside the  region surrounded by CMS
(in which the state in the r.h.s. is $lighter$ than the gluon)
 even a
notion of a gluon as a separate state does not exist, and using
the
``magnetic language'' (r.h.s.) for its description becomes mandatory.\\ \\

\subsection{Classical Molecular Dynamics for non-Abelian plasmas}

Another direction, pioneered by Gelman, Zahed and one of us
\cite{GSZ}, is to use experience of classical strongly coupled
electromagnetic plasma. Their model for the description of
strongly interacting quark and gluon quasiparticles  as a
classical and non-relativistic Non-Abelian Coulomb gas. The sign
and strength of the inter-particle interactions are fixed by the
scalar product of their classical {\it color vectors} subject to
Wong's equations. The EoM for the phase space coordinates follow
from the usual Poisson brackets: \be \{  x_{\alpha\,i}^m,
p_{\beta\,j}^n \}=\delta^{mn} \delta_{\alpha\beta}\delta_{ij}
\,\,\,\, \{ Q_{\alpha\,i}^a, Q_{\beta\,j}^b\}=
f^{abc}\,Q_{\alpha\,i}^c \ee For the color coordinates they are
classical analogue of the SU(N$_c$) color commutators, with $
f^{abc}$ the  structure constants of the color group. The
classical color vectors are all adjoint vectors with
$a=1...(N_c^2-1)$. For  the non-Abelian group SU(2) those are 3d
vectors on a unit sphere, for SU(3) there are 8 dimensions minus 2
Casimirs=6 d.o.f.\footnote{ Although color EoM does not look like
the usual canonical relations between coordinates and momenta,
they actually are pairs of conjugated variables, as can be shown
via the so called Darboux parametrization, see \cite{GSZ} for
details.}.

This cQGP model was studied using Molecular Dynamics (MD),
the equations of motion were
 solved numerically  for $n\sim 100$ particles.
It also displays a number of phases as the Coulomb coupling is
increased ranging from a gas, to a liquid, to a crystal with
anti-ferromagnetic-like color ordering. There is no place for details
here, let us only mention that important transport properties like
 diffusion and viscosity vs coupling.
note how different and nontrivial they are.
When extrapolated to the sQGP
suggest that the phase is liquid-like, with a diffusion constant
$D\approx 0.1/T$ and a bulk viscosity to entropy density ratio
$\eta/s\approx 1/3$.
The second paper of the same group\cite{GSZ} discussed the energy
and
the screening at $\Gamma>1$, finding large deviations from the Debye
theory.

The first study combining classical MD with quantum treatment
of the color degrees of freedom has been attempted by the Budapest
group \cite{Hartmann:2006nb}.

\subsection{Electric-magnetic dualities in supersymmetric theories}
Progress in supersymmetryc (SUSY)  Quantum Field Theories
 was originally stimulated
by a desire to get rid of perturbative divergencies and solve the
so called hierarchy problems. However in the last 2 decades it
went much further than just
 guesses of possible dynamics at superhigh energies.
 A fascinating array of nonperturbative phenomena have been
discovered in this context, making them into an excellent theoretical
 laboratory. However we think
that their relevance to QCD-like theories are
neither understood not explored in a sufficient depth yet.

Studies of instantons in these theories have resulted in exact beta
functions \cite{SVZ_beta} and other tools, which have allowed
Seiberg to get quite complete picture of the phase structure
of  \N=1 SUSY gauge theories \cite{Seiberg}.

This was enhanced in the context of  \N=2 SUSY gauge theories by
Seiberg and Witten \cite{SeiWit}, who were able to show how
physical content of the theory changes as a function of Higgs VeVs
(in a ``moduli space'' of possible vacua).
 Singularities in moduli space were identified with the phase transitions, in
which one of the MQPs gets massless. Seiberg and Witten have
found a fascinating set of $dualities$, explaining where and how a
transition from one language to another (e.g. from ``electric'' to
``magnetic'' to ``dyonic'' ones) can explain what is happening at
the corresponding part of the moduli space, in the simplest and
most natural way.

One lesson from those works, which is most important for us, is
what happens with the strength of electric $e$ and magnetic
coupling $g$ near the phase transition. As e.g. monopoles gets
light and even massless at some point, the ``Landau zero charge''
in the IR is enforced by the $U(1)$ beta function of the magnetic
QEDs, making them weakly coupled in IR, $g<<1$. The
  Dirac quantization  (\ref{Dirac_quantization}) therefore
demands that the electric coupling must get large $e>>1$,
enforcing the ``strongly coupled'' electric sector in this region.

Since two pillars of this argument -- $U(1)$ beta function and
Dirac quantization -- do not depend on supersymmetry or any other
details of the SW theory, we therefore now propose it to be a
generic phenomenon. We thus conjecture it to be also true near the
QCD deconfinement transition $T\approx T_c$, explaining why
phenomenologically we see a strong coupling regime there.

The high-$T$ limit, on the other hand, is similar to large-VEV
domain of moduli space: here the $SU(N)$ asymptotic freedom
in UV plus screening makes the electric charge small. Thus here
MQPs are heavy and strongly coupled.

\subsection{Lessons from lattice gauge theory}
 {\bf Static potentials.
} One of the principal reasons we proposed to change the
traditional viewpoint of putting confinement at the center, can be
explained using lattice data on the $T-$dependence of the so
called ``static potentials''. The traditional reasoning points to
the free energy $F(r,T)$ associated with static quark pair
separated by a distance $r$, and defines the deconfinement as the
disappearance of a (linearly) growing ``string'' term in it, so
that at $T>T_c$ there is a finite limit of the free energy at
large distances, $F(\infty,T)$. This phenomenon has often been
referred to as a ``melting of the confining string'' at $T_c$.

\begin{figure}[t]
  \epsfig{file=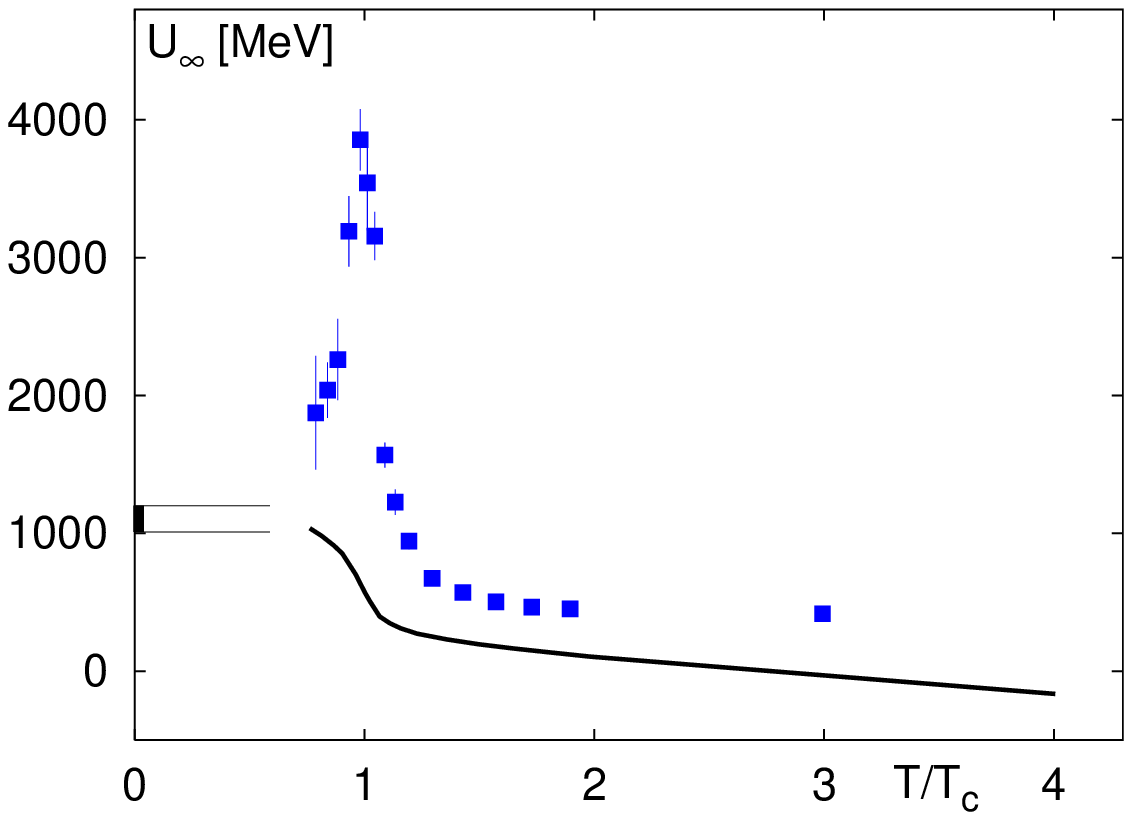,width=7.5cm}
  \epsfig{file=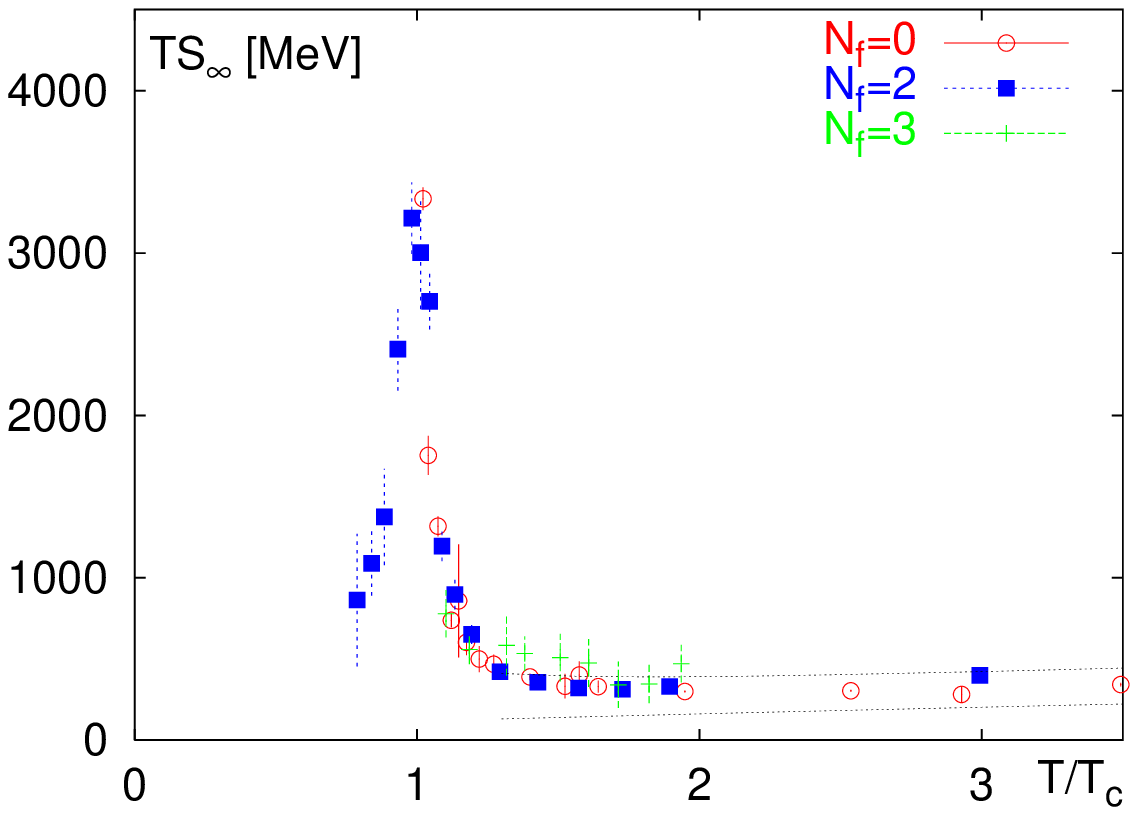,width=7.5cm}
\caption{
  The energy (a) and entropy (b)
(as $TS_\infty(T)$) derived from the free energy
of two static quarks separated by large distance,
in $2$-flavor QCD according to
 \protect\cite{Kaczmarek:2005zp}.
}
\label{fig_hq}
\end{figure}

However, as explained by Polyakov nearly 3 decades ago
\cite{Polyakov}, the string actually should not disappear at
$T_c$: at this point its energy gets instead $compensated$ by the
entropy term so that the $free$ energy $F=U-TS$ vanishes. As
detailed lattice studies revealed, in fact the energy and entropy
associated with a static quark pair are strongly peaked exactly at
$T\approx T_c$, see Fig.\ref{fig_hq}. The potential energy is
really huge there, reaching about 4 GeV(!), while the associated
entropy reaches equally impressive value of about 20. Nothing like
that can be explained on the basis of Debye-screened weakly
coupled gas of EQPs -- the usual picture of QGP until few years
ago. We think that the explanation of such large energy and huge
number $\sim exp(20)$ of occupied states can only be obtained if
several correlated quasiparticles are bound to heavy charges,
presumably in the form of gluonic chains or ``polymers''
\cite{Liao_1} conducting the electric flux from one charge to
another.

Therefore, the ``deconfinement'' seen in disappearing linear term in
 free energy
is actually restricted to static (or adiabatically slowly moving)
charges, while for finite-frequency
 motion of light  or even heavy (charmed) quarks
one still should find mesonic bound states even in the deconfined
phase \cite{SZ_bound}.
Lattice studied of light quark and  charmonium states \cite{charmonium}
 found that
they indeed persist till $T\approx 2T_c$: this conclusion was
dramatically confirmed by experimental discovery that $J/\psi$
suppression at RHIC is smaller than expected and is consistent
with a new view, that $J/\psi$ is $not$ melting at RHIC (where
$T<2T_c$).

One set of well-known lattice studies have tried to answer the
following questions: Is a ``dual superconductor'' picture
consistent with what is observed on the lattice? In particular, is
the shape and field distribution inside the confining strings in
agreement with that in the Abrikosov flux tube of a superconductor
(Abelian Higgs model)? As one can read e.g. in \cite{Balistrings},
the answer seems to be a definite yes. Can one define in some way
monopoles and their paths, and are those (in average) consistent
with dual Maxwell equations? As one can read in e.g.
 \cite{max_abelian_monop}, the answer seems to be also yes.

Unfortunately, those studies (as reviewed in \cite{Baker:1991bc})
 were mostly concentrated in the vacuum
 $T=0$, while we are interested by the
deconfined plasma $T>T_c$. Is there any general reason to think
that MQPs play an important role here as well? The most important
argument\footnote{Note a principle difference with all
electromagnetic plasmas, which have no magnetic screening at all.
For example, solar magnetic flux tubes are extended for a millions
of km, with unimpeded flux. } is the persistence of {\em static
magnetic screening} at all $T$, up to infinitely high $T$.

{\bf Screening}.  Although  static magnetic screening was shown to
be absent in perturbative diagrams \cite{Shur_QGP}, it has been
conjectured by Polyakov \cite{Polyakov} to appear
nonperturbatively at the ``magnetic scale'' which at high $T$ is
 \be \Lambda_M=e^2 T\ee
 The magnetic screening mass and monopole density should thus be
\be M_M =C_M \Lambda_M, \hspace{1cm} n_M=C_n  \Lambda_M^3 \ee
with some numerical constants $C_M, C_n$.

\begin{figure}
\hspace{+0cm} \centerline{\epsfxsize=8.cm\epsffile{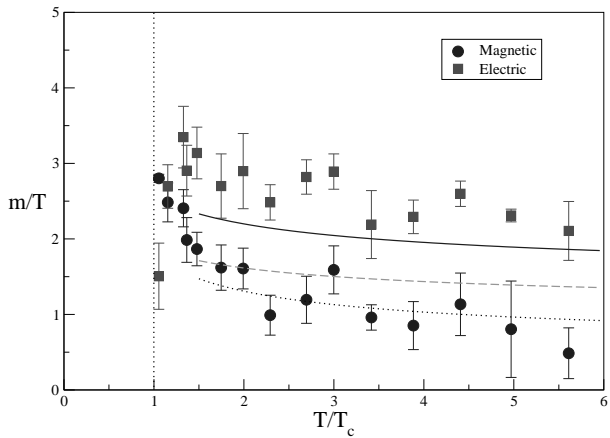}}
\hspace{+0.5cm}
 \caption{Temperature
dependence of electric and magnetic screening
  masses according to Nakamura et al \protect\cite{Nakamura:2003pu}.
The dotted line is fitted by the assumption, $m_g \sim g^2T$. For
the electric mass, the dashed and solid lines represent LOP and
HTL resummation results, respectively. \label{fig_emt}}
 \end{figure}

  To illustrate current lattice results, we
show the $T$-dependence of the electric and magnetic screening
masses calculated by Nakamura et al \cite{Nakamura:2003pu}, see
Fig.\ref{fig_emt}. Note that electric mass is larger than magnetic
one at high $T$, but vanishes at $T_c$ (because here electric
objects gets too heavy and effectively disappear). The magnetic
screening mass however grows toward $T_c$, which is consistent
with its scaling estimate \be M_M^2\sim (e^2 T)^2\ee (Another
estimate of the magnetic screening can be done in the dual
language as \be M_M^2\sim g^2 n_M/T\sim g^2 (e^2 T)^3/T\ee which
is
 a perturbative (small magnetic coupling
$g$) loop: note that it agrees with the former one due to Dirac
condition $e\sim 1/g$.)

If one uses screening masses to get an idea about density of
electric and magnetic objects, one finds that the point at which
electric and magnetic masses are equal should be close to the {\em
E-M equilibrium} point we emphasized above.
 This argument places the equilibrium temperature somewhere in the region
of \be T_{E=M}\approx (1.2-1.5)T_c=250-300\, MeV\ee.

{\bf High-T monopoles.} The total pressure related to magnetic
(3d) sector of the theory and especially the spatial string
tension are other observable related to MQPs above $T_c$: for a
short recent summary see  \cite{KorthalsAltes:2006gx}. Two
important points made by Korthals-Altes are: (i) MQPs must be in
the $adjoint$ color representation, to explain data on k-strings
and magnetic pressure; (ii) there seems to be a nontrivial small
``diluteness'' parameter of the MQPs ensemble \be \delta=
{\sigma_1\over M_M^2}\approx {(N-1) n_M \over
  M_M^3}\approx {1 \over 20} \ee
The fact that screening takes place at distances $smaller$ than
the average inter-MQP ones is a clear indication that screening is
not a Debye-type weak coupling one, but rather the opposite
$strongly$ coupled (correlated) screening\footnote{ If a reader
may have doubts that a correlated screening may produce such a
result, here is an example from the physics of the QCD instantons.
The typical inter-instanton distance $n^{-1/4}\sim 1\, fm$ is 5
times $larger$ than the screening length of the topological charge
$R_{top}=1/M(\eta')=.2\, fm$: the corresponding ratio for
monopoles seem to be around $\delta^{-1/3}\sim 3$. In both cases
we don't know how exactly the opposite charges are correlated:
pairs or chains are two obvious possibilities. }.

{\bf Dyons}. A very special sector of MQPs are particles with both
charges. Because they produce parallel electric and magnetic
fields, they have nonzero $(\vec E  \vec B)$ and thus the
topological charge. In fact, as shown by Kraan et al
\cite{Kraan:1998kp}, finite-$T$ instantons can be viewed as being
made of $N_c$ self-dual dyons: for a very nice
 AdS/CFT ``brane-based'' construction leading
to the same conclusion, see \cite{Lee-Yi}.

Topology is in turn associated with the Dirac zero eigenvalues for
fermions, which can be located and counted on the lattice quite
accurately. Furthermore, a ``visualization'' of dyons inside
lattice gauge field configurations (using variable non-trivial
holonomy) has been developed into a very sensitive tool
\cite{Gattringer:2006wq}, revealing multi-dyon configurations and
their dynamics. One can verify that they make a rather dilute but
highly correlated systems: in fact closed chains of up to 6 dyons
of alternating charges have been seen. The self-dual dyon density
and other properties, as well as their relation to instantons and
confinement are summarized in recent paper \cite{Gerhold:2006sk}.
It is enough to mention only that self-dual dyons, like
instantons, are electrically screened \cite{Shu_78,YaffePisarski}
and thus rapidly disappear into the QGP at $T>T_c$. Around $T_c$
their density can thus be related to the instanton density \be
n_{dyon}\sim N_c n_{instantons}/T\sim 3 fm^{-3}\ee and the mass to
the instanton action \be M_{dyon}=T*S_{instanton}/N_c\sim
(3-4)T\ee Both are of the order of the density (and the mass) of
the electric (gluon and quark) quasiparticles at $1.5T_c$,
confirming a suggested E-M equilibrium in this region.

\subsection{Higgs phenomenon in QGP?}

In this subsection we would like to comment, in a brief form, on a
number of questions which are invariably asked in connection with
Higgs phenomenon and monopoles at $T>T_c$.

Naively, there is no simple and direct way to apply the lessons
from supersymmetric theories such as $\cal N$=2 Seiberg-Witten
theory to QCD-like setting. The former has scalar fields and flat
``moduli space'' of possible vacua, while the latter has neither
scalars nor supersymmetry to keep the moduli space flat.

At finite $T$ the role of Higgs field is delegated to temporal
component $A_0$ of the gauge field: and in fact in gluodynamics
there is a spontaneous breaking of the $Z(N_c)$ symmetry at
$T>T_c$ because the corresponding effective action
$S_{eff}(<A_0>)$ has $N_c$ discrete degenerate minima\footnote{
Fermions will lift this degeneracy, as is well known.}.

Furthermore, the corresponding effective action gets
small near $T_c$ and large fluctuations in ``Higgs VEV''
 $<A_0>$ are seen in lattice configurations; so one may
think first about a generic case in which it is some (color matrix
valued) constant in each configuration, to be averaged with
appropriate weight  $exp[-S_{eff}(<A_0>)]$ later. Thus one may
think about an explicit adjoint Higgs breaking of the color group,
parameterized by $N_c-1$  real VEVs (e.g. for $SU(3)$
$Tr<A_0\lambda^a>$with Gell-Mann diagonal matrices a=3,8). Such
breaking makes all gluons massive, except the remaining unbroken
$N_c-1$ U(1)'s which remain massless. These remaining U(1)'s are
the Abelian gauge fields which define magnetic charges of the
monopoles and their long-range interactions (and electric ones, in
the case of dyons).

Finally, the last comment about one lesson from SUSY theories
which we don't think can be transferred into the QCD world: these
are the enforced properties of monopoles (and many other
topological objects like branes) which happen to be ``BPS states''
with their Coulomb interactions being exactly cancelled by
massless scalar exchanges. As a result, such objects can often
``levitate'' in SUSY settings. In QCD we however do not see or
need massless scalars, leaving the usual Coulomb and Lorentz
forces dominant at large distances.

\section{Few-body problems with Magnetic charges }

The simplest few-body system with magnetic charge is made of two
particles: one has electric charge and the other has magnetic
charge. In a more general sense we should consider them as two
dyons, both with nonzero electric as well as magnetic charges.
This problem has been very well studied for many years in both
classical and quantum mechanics, and it has fascinated the
physicists with many unusual features. See for example
\cite{Jackson}\cite{Goldhaber}\cite{Milton}.

In such problems one has  both electric and magnetic
fields. We have the electric field from an E-charge
(at space point $\vec r_e$) to be
\begin{equation}
\vec E (\vec r) =e \frac{\vec r-\vec r_e}{|\vec r - \vec r_e|^3}
\end{equation}
and magnetic field from a M-charge (at space point $\vec r_g$) to be
\begin{equation}
\vec B (\vec r) =g \frac{\vec r - \vec r_g}{|\vec r - \vec r_g|^3}
\end{equation}
The interaction between
 one moving dyon $(e_1,g_1)$ and the other $(e_2,g_2)$ is
given by the Coulomb and $O(v/c)$ Lorentz forces
\begin{eqnarray} \label{dyon_force}
\vec F_{12} =&& (e_2\cdot e_1 + g_2 \cdot g_1) \frac{\vec r}{r^3}
\nonumber \\
&&+ (e_2\cdot g_1 - g_2 \cdot e_1 ) \frac{\vec{v}_2}{c} \times
\frac{\vec r}{r^3} +O(v^2/c^2)
\end{eqnarray}
with $\vec r = \vec r_2 - \vec r_1$. Here we have used the
Gaussian units in which $\vec E$ and $\vec B$ have the same unit
and so are the charges $e$ and $g$.

 As early as in 1904, J. J. Thomson found the
even two non-moving charges
have a nonzero angular momentum $J$
carried by rotating {\it electromagnetic field }.
Indeed, for  an E-charge and a M-charge (separated by $\vec r$) as
sources, it is
\begin{equation} \label{Thomson}
\vec J_{field} = \int d^3 x \vec x \times \frac{\vec E \times \vec B}{4 \pi c} = \frac{eg}{c} \hat r
\end{equation}
This angular momentum depends only on the value of charges,
 independent on how far or close they may be. Its direction is
radial,  pointing from the E-charge to the M-charge.

\begin{figure}[t]
  \epsfig{file=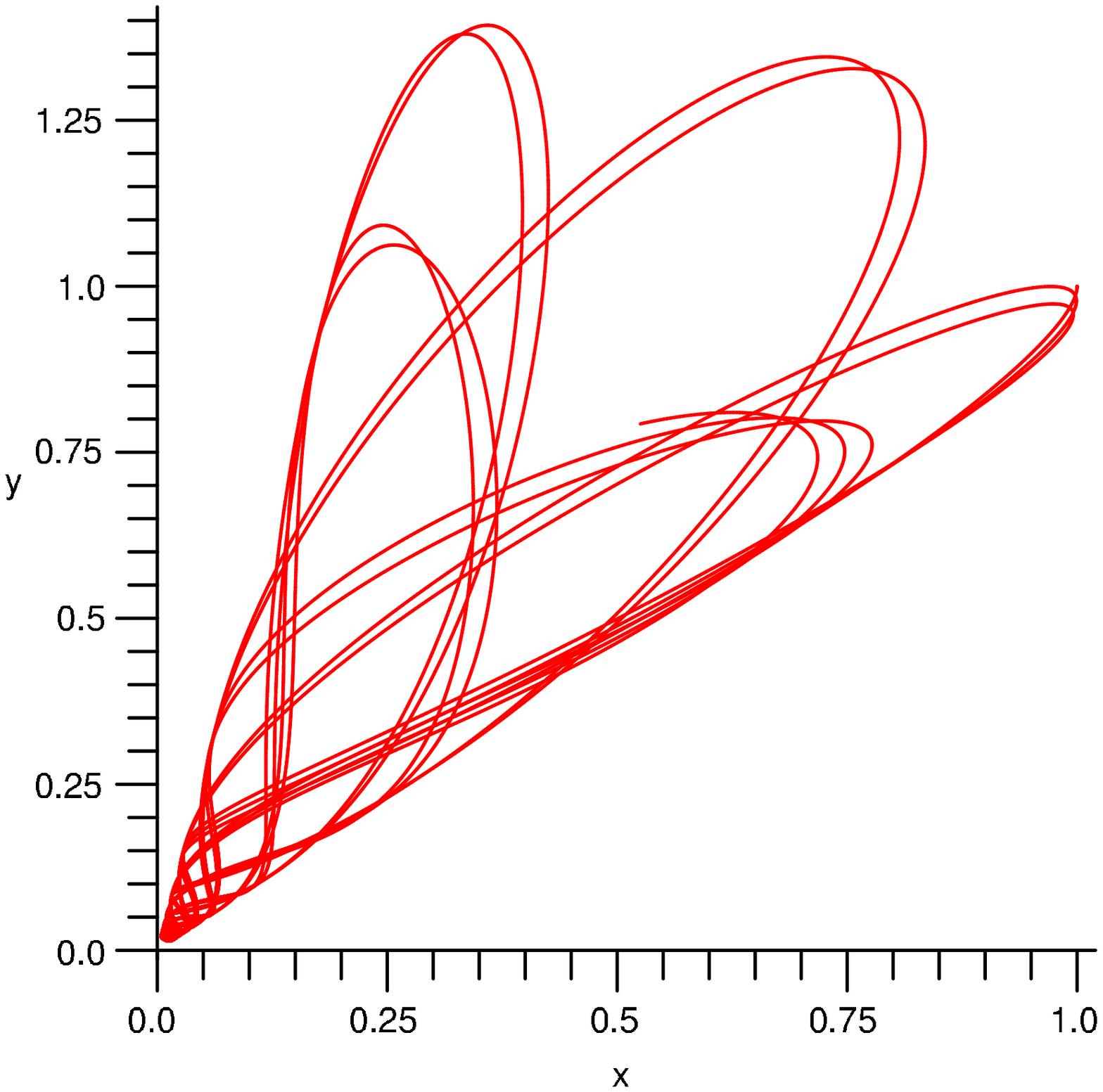,width=8.cm}
\caption{
  The trajectory of a dyon in a field of static charge.
}
\label{fig_cone}
\end{figure}

Even earlier, in 1896 Poincare observed that a dyon moves
in a charge Coulomb field on the surface of a
 cone, as shown in Fig.\ref{fig_cone}.
 Their relative motion (angular rotation and radial bouncing ) is
 always confined inside a cone simply because of the conservation of
 total angular momentum including the relative rotation and the
 field's angular momentum (\ref{Thomson}) as well. When getting closer
 to each other the two particles are forced to rotating faster thus
 experiencing effective repelling which makes them bouncing radially.
Another way to explain the conical motion is to notice that
a magnetic charge is making Larmor circles around the electric field;
it shrinks near the charge because the fields gets stronger there.

 Quantum mechanics of such two-dyon system has been worked out in
 many details since 1970s, especially the many bound states are
 calculated, see \cite{Goldhaber,Milton} for review.

\subsection{Static Electric Dipole and a Dynamical Monopole}
A very interesting and important few-body problem is a magnetic
monopole moving in the field of a static electric dipole. This is
a starting point for studying a "color"-electric
dipole(quark-anti-quark) surrounded by a gas of weakly interacting
monopoles which, as we argue, may be very much relevant for
understanding confinement. Also as far as we know, this system
seems never been 
studied before.

\begin{figure}
\hspace{+0cm}
\centerline{\epsfxsize=9.cm\epsffile{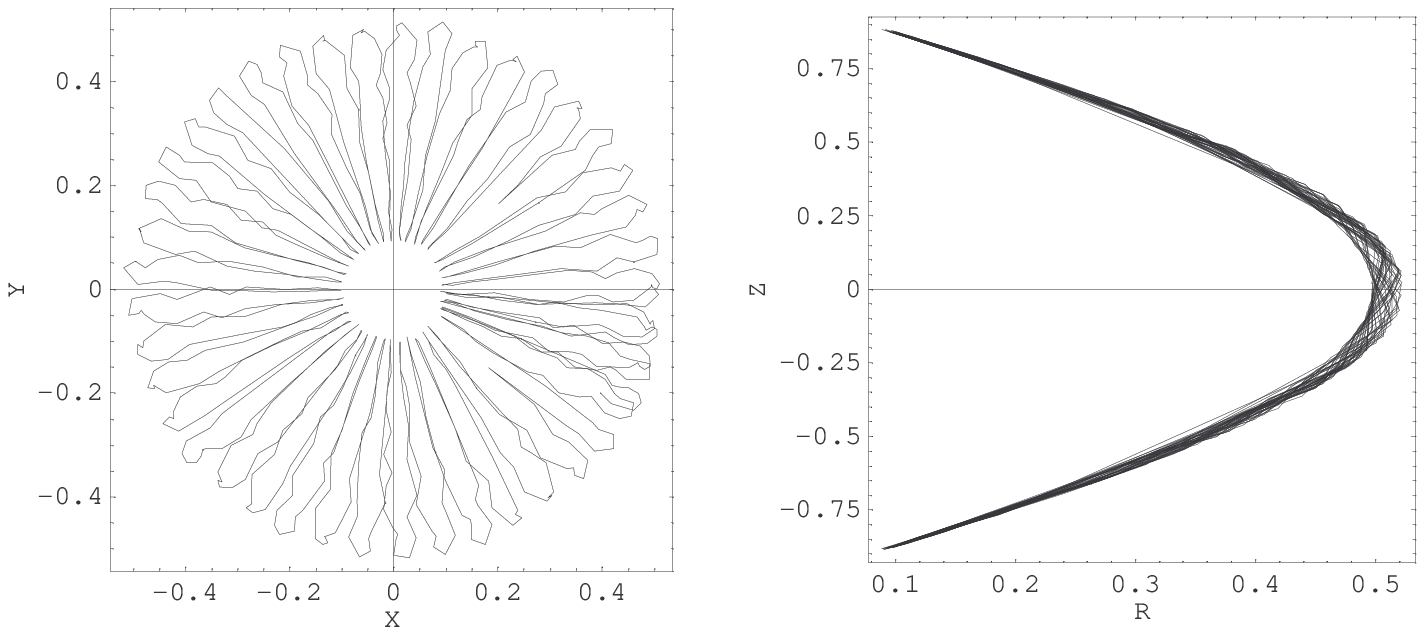}}
\hspace{+0.5cm}
 \caption{\label{fig_trap_1}
Trajectory of monopole motion in a static electric dipole field
(with charges at $\pm 1 \, \hat z$) as (left panel)projected on
x-y plane and (right panel)projected on R-z plane
($R=\sqrt{x^2+y^2}$).}
 \end{figure}

\begin{figure}
\hspace{+0cm}
\centerline{\epsfxsize=9.cm\epsffile{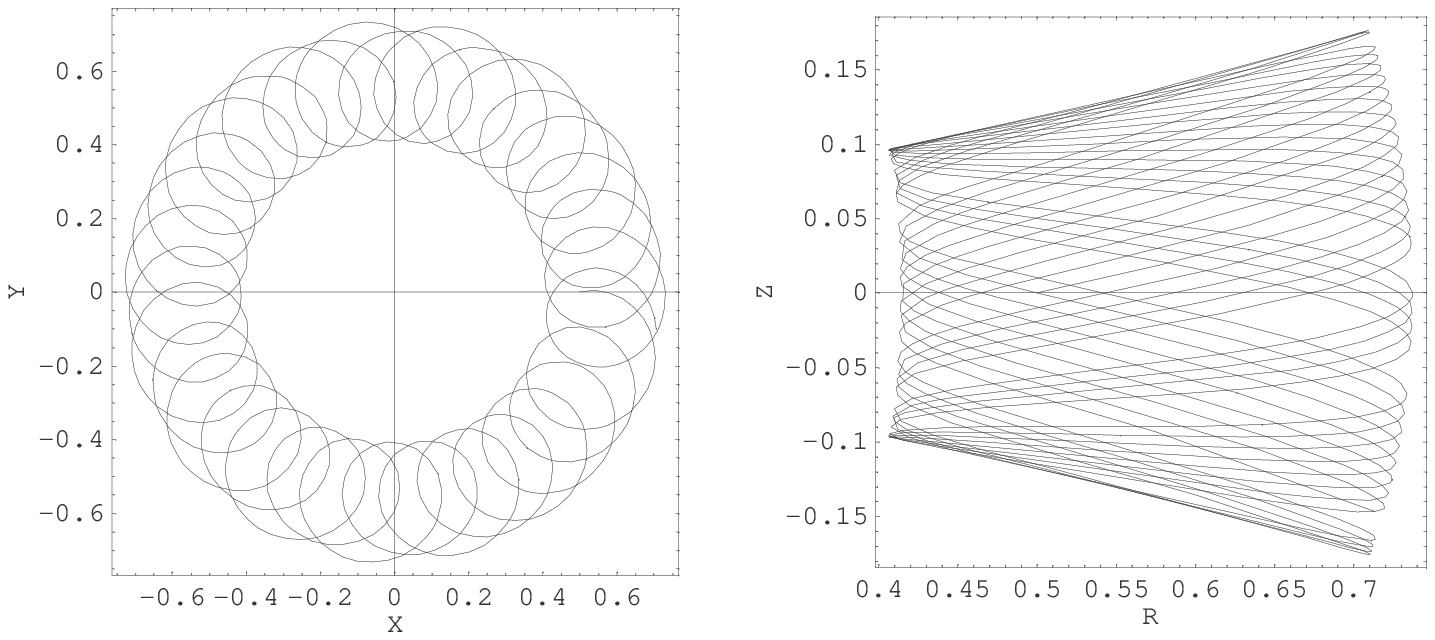}}
\hspace{+0.5cm}
 \caption{\label{fig_trap_2}
Trajectory of monopole motion in a static electric dipole field
(with charges at $\pm 1 \, \hat z$) as (left panel)projected on
x-y plane and (right panel)projected on R-z plane
($R=\sqrt{x^2+y^2}$). }
 \end{figure}

Considering quark-anti-quark pair surrounded by a gas of weakly
interacting monopoles as the scenario for confining the flux tube
just around $T_c$, one realize that to find possible bound states
(namely, states with the monopole attached around the electric
dipole permanently or at least for long time before "decaying"
away) of such a system is potentially a key to understand the
source of large entropy associated with static quark-anti-quark as
indicated by lattice. We will discuss this issue in both classical
and quantum mechanics.

In classical mechanics, we have the EoM for the monopole (of mass
$m$ and magnetic charge $g$) to be the following:
\begin{equation} \label{monopole_dipole_classical}
m \frac{d^2 \vec r}{dt^2} = g \vec E_d \times \frac{d\vec r}{dt}
\end{equation}
with $\vec E_d$ the electrostatic field from the dipole with $\pm
e$ charges sitting at $\pm a$ on $z$ axis:
\begin{equation} \label{dipole_field}
\vec E_d = e \,  { \bigg[ }\frac{\vec r-a\hat z}{|\vec r - a \hat
z|^3} - \frac{\vec r+a\hat z}{|\vec r + a \hat z|^3} {\bigg ]}
\end{equation}
It will be much convenient to work in the cylinder coordinate
$(\rho,\phi,z)$. By running this EoM numerically with various
initial conditions, we can directly obtain real time trajectories
of the monopole.

\begin{figure}
\hspace{+0cm}
\centerline{\epsfxsize=9.cm\epsffile{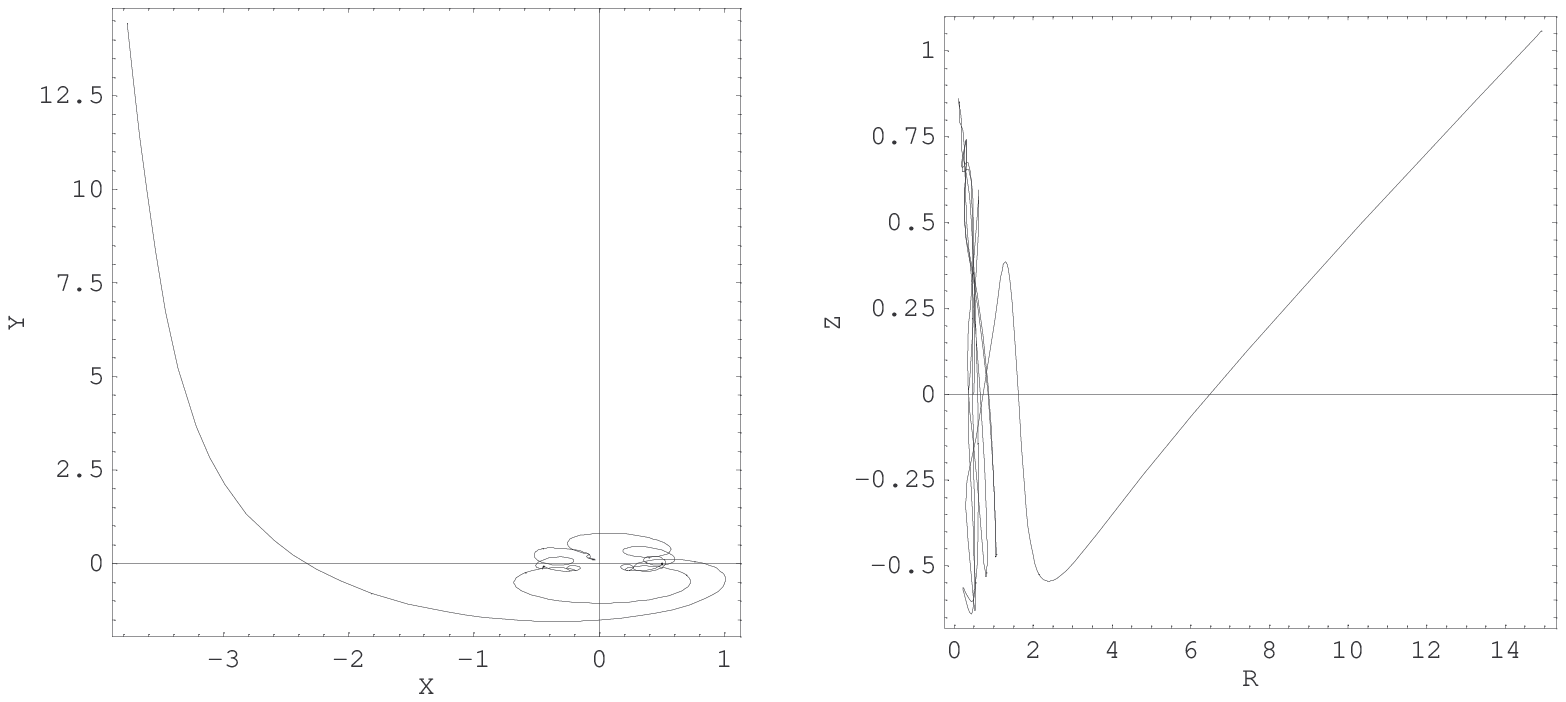}}
\hspace{+0.5cm}
 \caption{\label{fig_escape}
Trajectory of monopole motion in a static electric dipole field
(with charges at $\pm 1 \, \hat z$) as (left panel)projected on
x-y plane and (right panel)projected on R-z plane
($R=\sqrt{x^2+y^2}$). }
 \end{figure}

A lot of very complicated and very different motions have been
found, sensitively depending on the initial conditions. Roughly
one may divide these trajectories into two categories: "trapping"
cases (see Fig.\ref{fig_trap_1} and Fig.\ref{fig_trap_2}) and
"escaping" cases (see Fig.\ref{fig_escape}). By "trapping" cases
we mean the monopole starts with $|\vec r| \sim a$ and after a
relatively long time it still remains within distance $\sim a$
from the dipole, while in "escaping" cases the monopole begins
moving further and further away from the dipole after a somewhat
short time. Due to limited space we show below only few pictures
for both cases. Let's just emphasize one particular feature as
clearly revealed in Fig.\ref{fig_trap_1}: the monopole is bouncing
back and forth between the two electric charges, because of the
effective repulsion when it is getting close to the charges (as
has been explained in the charge-monopole motion). We have found
many such cases which look like two standing charges playing E-M
"ping-pong" with the monopole. So there are many classical bound
states for such system, and in principle one can scan through the
phase space of monopole's initial position and momentum to
estimate the "trapping" states' phase space volume.

This phenomenon as shown here is dual to the famous ``magnetic
bottle'', a device invented for containment of hot electromagnetic
plasmas, provided magnetic coils at its ends are substituting the
electric dipole and a moving monopole replaced by the electric
charge.

\begin{figure}
\hspace{+0cm} \centerline{\epsfxsize=8.cm\epsffile{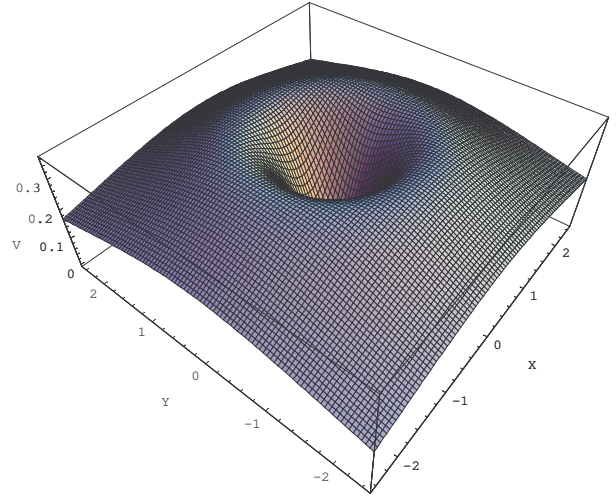}}
\hspace{+0.5cm}
 \caption{\label{fig_effective_pot}
 (color online) Quantum mechanical effective potential for a
 monopole in a static electric dipole field. See text for details.
}
 \end{figure}

Now let's turn to the quantum mechanics of such system. One can
write down the following Hamiltonian for the monopole:
\begin{equation} \label{dipole_hamiltonian}
\hat{\cal H} =\frac{(\vec p + g {\vec A_e})^2}{2m}
\end{equation}
Here $\vec A_e$ is the electric vector potential of the dipole
electric field, which can be thought of as a dual to the normal
magnetic vector potential of a magnetic dipole made of
monopole-anti-monopole. By symmetry argument we can require the
vector potential as $\vec A_e = A^{\phi}_e(\rho,z) \hat \phi$ and
the monopole wavefunction as $\Psi=\psi(\rho,z) e^{i f \phi}$ with
$f$ the z-angular-momentum quantum number. Then the stationary
Schroedinger equation is simplified to be
\begin{eqnarray} \label{effective_potential}
&&[\frac{\vec p_{\rho}^{\,\, 2}+ \vec p_z^{\,\, 2}}{2m} + V_{eff}
] \psi = E \psi
\\
&&V_{eff} = \frac{\hbar^2}{2m} [\frac{1}{\rho /a }
(\frac{ge}{\hbar}\frac{\rho A_e^{\phi}}{e}+f)]^2
\end{eqnarray}
To go further one has to specify a gauge (which is equivalent to
choosing some particular dual Dirac strings for the charges) so as
to explicitly write down $A_e^{\phi}$. We use the gauge which
corresponds to the situation with one Dirac string going from the
positive charge along positive $\hat z$ axis to $+ \infty$ and the
other going from the negative charge along negative $\hat z$ axis
to $- \infty$. This gives us:
\begin{equation} \label{vector_potential}
A_e^{\phi} =
-\frac{e}{\rho}[2+\frac{z-a}{\sqrt{\rho^2+(z-a)^2}}-\frac{z+a}{\sqrt{\rho^2+(z+a)^2}}]
\end{equation}
To give an idea of the effective potential we show
Fig.\ref{fig_effective_pot} where
$V_{eff}(\rho,z,f)=V_{eff}(\sqrt{x^2+y^2},z,f)$ is plotted for the
x-y plane with $z=0$ and $f=0$. From the plot we can see that
there must also be quantum states with the monopole bounded within
the potential well around the dipole for a long time before
eventually decaying away. Namely one can find states with
$E=\hbar \omega + i \Gamma$ with $\Gamma << \hbar \omega$ and count how many
there are. This issue will be separately addressed in details elsewhere.

\subsection{Flux tubes in a classical gas}
  The problem discussed in the previous subsection has of course
not only ``trapped'' monopole states emphasized above, but also a
lot of scattering states for monopoles with positive energy. In a
simple classical setting it is quite clear that in an ensemble
(gas) of monopoles, particles would scatter off the effective
positive potential depicted in Fig.\ref{fig_effective_pot}
 and provide effective pressure on it. The net result
should be nothing but stabilization of the $electric$ flux
tubes in a gas of monopoles, at a size equilibrating gas pressure
with the field stress tensor.

  The situation dual to the problem discussed is $magnetic$
flux tubes in a gas of electric charges. Since this does not need
magnetic charges, it takes place in electrodynamic plasmas. As a
well known example, stabilized magnetic flux tubes are found in
the solar plasma: there they extend for millions of kilometers and
are seen in a modern telescope as a substructure of ``solar black
spots''.

Although we would discuss classical flux tubes in a separate
paper, we would like to point out now their role in our general
picture. We believe those occur at two extremes: (i) at $T$ just
above $T_c$, as classical electric flux tubes are expelled by
dominant MQPs, and (ii) at very high $T$ when there are magnetic
flux tubes expelled by dominant EQPs. In the former case electric
flux tubes are a natural continuation of $quantum$ electric flux
tubes which are created by Bose-condensed MQPs in the confining
phase at $T<T_c$. What we would like to point out here is that
MQPs expel the electric field lines out of the volume they occupy
{\em irrespective whether they are Bose-condensed or not}.

\section{Molecular dynamics without periodic boxes}

Molecular dynamics (MD) provides a straightforward way to study
the various dynamical properties of a classical many-body system.
The system we are interested in is a plasma containing both
electric and magnetic (and both positive and negative) charges. So
in the normal convention used by plasma physics community, this is
a four-component-plasma(FCP). We however would rather name it as
2E-2M-plasma to explicitly show its content. More widely speaking
we may even include one more type of particles, namely dyons with
both electric and magnetic charges for individual particles,
making a 2E-2M-4D-plasma. In this paper we will report our results
for 2E-2M-plasma with three different contents: {\it pure electric
(which reduces to normal TCP) plasma, plasma with about one
quarter of particles as magnetic charges, and plasma with about
half of particles as magnetic charges, labelled throughout this
paper as M00, M25, M50 respectively}. Comparison among them is
expected to give indications about the role of magnetic charges,
especially in the transport properties. The microscopic dynamics
is classical EM, given by Newton's second law together with
electric Coulomb force (between two E-charges), magnetic Coulomb
force (between two M-charges) and $O(v)$ Lorentz force (between one
E-charge and one M-charge).

The routine MD method (as used in GSZ and most MD study of usual
plasma) is to put desired number of particles in a cubic box and
then include as many periodical image boxes (in all three
directions) as allowed by computing capacity. The summation over
images is very much time consuming especially for cases with
Coulomb type long range forces. Also energy conservation is not
very well preserved after long-time run because of the "kick" on
particles leaving one truncation boundary and entering
periodically on the opposite boundary which will gradually heat up
the system.

Here it should be emphasized that we have used an alternative
approach without any periodic boxes. What we have done is to
simply give all particles certain initial conditions and then let
them go. It turns out there are two different regimes which we
deal with separately: 1) "plasma in cup" at medium/weak coupling
regime, in which case we place a sharply rising large radial
potential barrier at certain radial distance to hold the particles
inside this "cup"; 2) "self-holding drop" at very strongly coupled
regime, which means the particles don't fall apart into small
pieces but behave like a little raindrop and so there is no need
for a "cup". In this way we are able to perform MD easily with
thousand particles and can conserve energy for less than percent
even after really long-time run. We will give more technical
details about our simulations in the second subsection while
present basic formulae, units and physical parameters in the first
subsection .

\subsection{Formulae,units and physical parameters}
For our 2E-2M-plasma, each particle has either electric charge or
magnetic charge. The E-charges are assigned as $e_i e$ with $e_i$
randomly and equally given $\pm 1$ ($e_i=0$ for M-charges) and the
M-charges are assigned as $g_i g$ with $g_i$ randomly and equally
given $\pm 1$ ($g_i=0$ for E-charges) too. For a pair of particles
their mutual force involves three combinations of their charges:
$e_{ij}=e_i \cdot e_j$, $g_{ij}=g_i \cdot g_j$, and an important
new one $\kappa_{ij}=e_i \cdot g_j - g_i \cdot e_j$. In present
study we use the same mass $m$ for both types of charges.

The equation of motion for the $i$th component particle is given
by:
\begin{eqnarray} \label{MD_EoM}
m \frac{d^2 \vec r_i}{dt^2} = \sum_{j \ne i} {\bigg [}
\frac{C}{r_{ij}^{K+1}} \hat{r}_{ji} &&+ \frac{e^2 \, e_{ij} }{r_{ij}^2} \hat{r}_{ji}  \nonumber \\
&& + \frac{g^2 \, g_{ij}}{r_{ij}^2} \hat{r}_{ji} + \frac{ge \,
\kappa_{ij}}{r_{ij}^2} \frac{d \vec r_i}{c \, dt} \times
\hat{r}_{ji} {\bigg ]}
\end{eqnarray}
where $\vec r_{ji}=\vec r_i - \vec r_j$. The first term on RHS is
the well-known necessary repulsive core without which all
classical plasma will collapse sooner or later since no quantum
effect arises at small distance to prevent positive charges
falling onto negative partners. We choose $K=9$ in our MD, which
is the same as some previous work\cite{GSZ}\cite{Hansen}. There is
no particular meaning for $K=9$ except that we want a large value
of $n$ which leads to relatively small correction ($\sim 1/K$) to
potential energy between $+e$ and $-e$ at and beyond the
equilibrium distance.

To set the units in our numerical study, we use the following
scaling for length and time (the unit of mass is naturally set by particle
mass $m$)
\begin{eqnarray} \label{unit_scaling}
&&  \vec{ \tilde{r}} = \vec r / r_0 \qquad with \qquad r_0=(C/e^2)^{\frac{1}{K-1}} \nonumber \\
&& \tilde t = t / \tau \qquad with \qquad
\tau=(m r_0^3/e^2)^{\frac{1}{2}}
\end{eqnarray}
which leads to the dimensionless equation of motion
\begin{eqnarray} \label{EoM_scaling}
\frac{ d^2 \vec{\tilde{r}}_i}{d{\tilde{t}}^2 }= \sum_{j \ne i}
 {\bigg [} &&
\frac{1}{\tilde{r}_{ij}^{n+1}} \hat{r}_{ji} + \frac{e_{ij} }{\tilde{r}_{ij}^2} \hat{r}_{ji}  \nonumber \\
&& + (\frac{g}{e})^2 \frac{g_{ij}}{\tilde{r}_{ij}^2} \hat{r}_{ji}
+ (\frac{g}{e} \frac{r_0/\tau}{c})
\frac{\kappa_{ij}}{\tilde{r}_{ij}^2} \frac{d \vec{\tilde{r}}_i}{d
\tilde{t}} \times \hat{r}_{ji} {\bigg ]}
\end{eqnarray}
With these setting, we have for example: Length = \# $\times
r_0$, Time = \# $\times \tau$, Frequency = \# $\times
\frac{1}{\tau}$, Velocity = \# $\times \frac{r_0}{\tau}$,
Energy = \# $\times \frac{e^2}{r_0}$, etc. All numbers
obtained from numerics are subjected to association with proper
dimensional factors in our units.

Now we still have two dimensionless physical parameters which controls in the
above the magnetic-related coupling strength:\\
1. $\tilde{g}=\frac{g}{e}$ : this parameter characterizes the
relative coupling strength of magnetic to electric sector. In
principle there is no limitation for it from classical physics.
Since we want to focus on the parametric regime which may be
relevant to sQGP problem near $T_c$ (where electric sector gets
strongly coupled while magnetic sector gets weak), the parameter
$\tilde{g}$ is expected to be small, so we will use
$\tilde{g}=0.1$ in the MD calculation. Naively suppose one has a
quantum problem with same $\tilde{g}=g/e$, then by combination
with the minimum Dirac condition $\frac{eg}{\hbar c} =
\frac{e^2}{\hbar c} \tilde{g}  = \frac{1}{2}$, one gets
$\alpha=\frac{e^2}{\hbar c}= 1/(2\tilde{g})\sim 5$ which is indeed
very strongly coupled.
 \\
2. $\beta=\frac{r_0 / \tau}{c}=\sqrt{\frac{e^2/r_0}{m
c^2}}$ : this parameter tells us how relativistic the particles'
motion will typically be. The importance of this parameter lies in
that it controls the strength of Lorentz force ($\beta \cdot
\tilde{g}$) between E-M charges. An important observation here is
that compared to the Lorentz coupling of a pure electric plasma
(which is $\beta^2$ from electric current-current) our Lorentz
force has only the first power of the small parameter $\beta$ and
is thus enhanced because of the existence of magnetic charges.
Since we are doing non-relativistic molecular dynamics, a small
value of $\beta$ should be chosen. In the sQGP the typical speed
is estimated to be about $1/3$ of $c$, in present calculation we
however will choose $\beta=0.1$ which on one hand is not far from
$1/3$ and yet on the other hand limits the relativistic
corrections to be not more than few percent.

One more physical parameter we should mention here is the so
called plasma parameter $\Gamma$ defined as the ratio of average
potential kinetic energy (neglecting the sign)
\begin{equation} \label{gamma_parameter}
\Gamma= {\bigg | } \frac{<U>}{<E_k>} {\bigg | } =  {\bigg | } \frac{<U/N>}{3k_B T /2} {\bigg | }
\end{equation}
This definition looks a little different from others\cite{GSZ}\cite{Hansen} where usual MD study with periodic boxes defines $\Gamma=\frac{e^2/a}{k_B T}$ with $a=(3/4\pi n)^{1/3}$. They use this conveniently because in their approach the density is fixed as desired, while in our case there is no boxes any more and we use the direct ratio which is essentially meaning the same thing. The difference and relation between the two $\Gamma$ values will be further discussed in section \ref{mapping}.

The plasma parameter $\Gamma$ is important in that:\\
1. it distinguishes strongly coupled plasma $\Gamma>>1$ and weakly coupled plasma $\Gamma<<1$;\\
2. it roughly distinguishes a gas phase $\Gamma<1$, a liquid phase
$\Gamma \sim 10$ and a solid (or solid-like) phase $\Gamma > 100$ ;\\
3. different types of plasma with the same value of $\Gamma$ could be compared in order to reveal the dependence of macroscopic properties on plasma contents and microscopic dynamics, and so we will measure properties as a function of $\Gamma$.

\subsection{Details of MD simulations }

In our MD simulations, 1000 particles are initially placed on the
sites of a $10\times 10\times 10$ cubic lattice with lattice
spacing $a=1.2 r_0$ \footnote{This value is very close to
$a=1.18 r_0$ which is calculated to be the equilibrium value of
$NaCl$-like structure under our repulsive core.}. They are given
electric charges $\pm 1$ in an alternating way in all 3
directions. Then for the M25 (M50) plasma, we randomly pick out
25\% (50\%) of the particles and re-assign them magnetic charges
instead of electric charges. Then all the particles are randomly
given initial velocity (for each of the three component) $
v^i_{1,2,3}(t=0)=V
* (RANDOM \#)$ ($RANDOM \#$ is between [0,1]) under condition that
the total velocity of the whole system is zero. For each type of
plasma, changing the value of $V$ can eventually lead to different
equilibrated system after certain time. Roughly the larger $V$ is,
the smaller (lower) the plasma parameter $\Gamma$ (temperature
$T$) will be. The total running time is $2000\Delta t$ with
$\Delta t = 0.1\tau$ (which is actually our output time step). In
general it takes $t_{therm} \sim 20-30 \tau$ to equilibrate the
system and we start measurements at $t=50\tau$. The iteration step
and accuracy in EoM subroutine is so chosen that the energy could
be conserved to less than few percent at the end of run. As
mentioned before, we have two different regimes which we will
discuss the details separately in the following.

\subsubsection{Plasma in cup}
Numerically we found that for about $\Gamma<25$, the little drop
we created couldn't hold itself and after some running time it
will break into a few much smaller pieces, which means the surface
tension is not large enough to maintain the original "big" drop.
To confine the particles in a finite volume and make them mix up
sufficiently, we put a radial potential barrier at some
cut-distance $R_{cut}$ to make a container holding such plasma:
\begin{equation}
V=[B*(r-R_{cut})]^L*\theta (r-R_{cut})
\end{equation}
By choosing $B=5$ and $L=11$ we make the edge of our "cup" a
really steep one, thus keeping as many particles inside the "cup"
as possible at all time because only very energetic particles are
able to climb up the edge a little and will soon be reflected
back. In our simulations we have used $R_{cut}=11 r_0$. In real
time of course the number of particles confined within the
$R_{cut}$ is always fluctuating, so are other macroscopic
quantities like energy etc. So this system is like a
grand-canonical ensemble. For this plasma in cup, all the
measurements are made for particles inside the cup only (namely
with $r\le R_{cut}$).

\begin{figure}
\hspace{+1.cm} \centerline{\epsfxsize=8.cm\epsffile{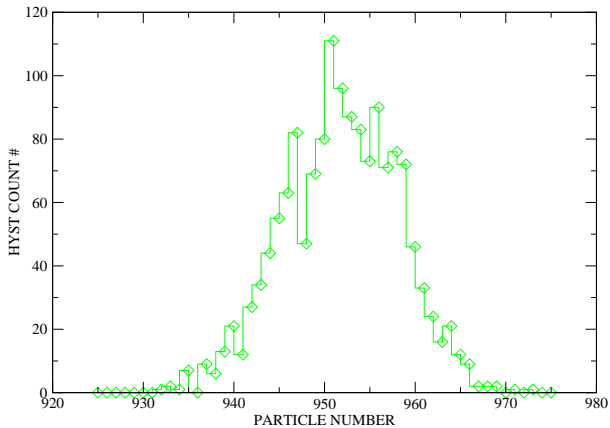}}
\hspace{+0.5cm}
 \caption{\label{fig_ct_fluct}
(color online) Histogram of total number of particles inside $R_{cut}$ at
1500 different time points. This is an example from M25 plasma at $\Gamma=0.99$.}
 \end{figure}

\begin{figure}
\hspace{+0cm} \centerline{\epsfxsize=8.cm\epsffile{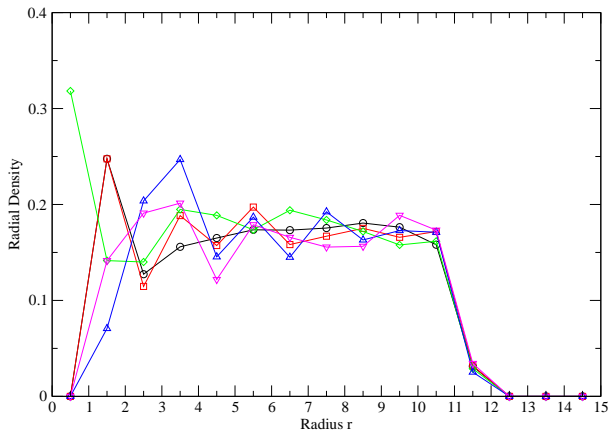}}
\hspace{+0.5cm}
 \caption{\label{fig_density}
(color online) Radial density of particles as a function of radius $r$. The
five curves are taken from different time points. This is an example from M25 plasma at $\Gamma=0.99$.}
 \end{figure}

By looking at the histogram of total number of particles at
different time points in Fig.\ref{fig_ct_fluct}, one see that the
system has very good distribution with well-defined average $N
\sim 950$ and $\sqrt{N}\sim 30$ fluctuation width. Another
important check is to see if the system is really homogeneous. In
Fig.\ref{fig_density} the radial local density $n (r)$ at five
different time points (from early to very late time) are shown,
from which it is clear that the density distribution is
homogeneous and stable enough. The large fluctuation at very small
r is understandable because one has much less particle numbers
$n 4\pi r^2$ for small r. One can also see that near our
cutting edge ($R_{cut}=11$) the particle density quickly drops
down as we want. These observations are true in all of our runs,
and the number density of our cupped plasma at different $\Gamma$
is controlled all at $n \approx 0.17$ with negligible variation. These
have shown that our simulations for cupped plasma is reliable.\\


\begin{figure}
\hspace{+0cm} \centerline{\epsfxsize=8.cm\epsffile{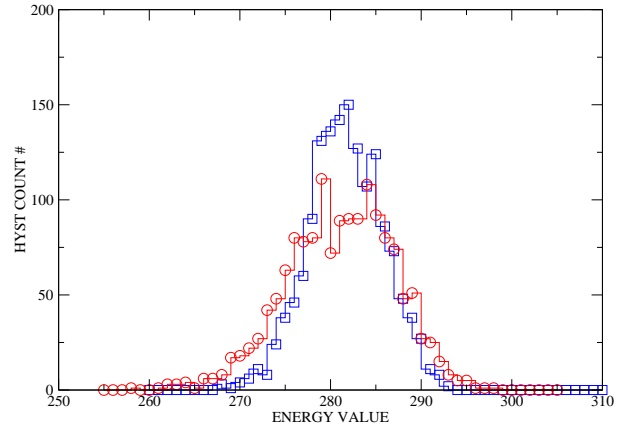}}
\hspace{+0.5cm}
 \caption{\label{fig_e_fluct}
(color online) Histogram of total kinetic(blue square) and potential(red circle)energy inside $R_{cut}$ at 1500 different time points. This is an example
from M25 plasma at $\Gamma=0.99$.
}
 \end{figure}

It is also important to check the fluctuation in energy. In
Fig.\ref{fig_e_fluct} we show a typical histogram of fluctuation
in kinetic and potential energy at all time points. Clearly both
distributions make complete sense and so are other macroscopic
variables which we skip because of limited space. Again these
justify our "plasma-in-cup" approach.\\

The results to be reported in Section IV. and V. are all obtained with
this method, which cover the $\Gamma$ value about $0.3-14$. We want to
focus on this region because it is most relevant to the sQGP.

\subsubsection{Self-holding drop}
For about $\Gamma>25$ we have found our little drop can,
amazingly, hold itself despite the possible expansion and
shrinking with considerable amplitude. By mapping the particles'
coordinates at the end of run we found the particles more or less
staying around their original positions. This very strongly
coupled system behaves more like a crystal, especially for $\Gamma
\to 100$. In this regime, we have found very good collective modes
which are shown to manifest themselves in the dynamical
correlation functions in a profound way. These results will be
reported in section VI. It should be pointed out that the
self-holding region is reached only for pure electric plasma (our
M00 plasma). For our M25/M50 plasma, with present method the
largest $\Gamma$ that can be achieved (after "cooling" and
equilibrating scheme) and maintained in a stable way is up to
$\sim 25$. The "cooling" method, namely turning on a braking force
proportional to particle velocity for some time and then turning
it off, can bring the M25/M50 system down to some instant $\Gamma
\sim 1000$ but then the system kinetic energy slowly but steadily
keeps increasing with potential energy getting more negative, the
overall effect of which eventually increases $\Gamma$ back down to
few tens. It seems indicating the mixture plasma refuses to become
solidified even at classical level because of Lorentz type force
(different from permanent liquid Helium which is due to quantum
effect). We will leave this issue for future investigation.

\section{Equation of state}

Before showing the data, we once again emphasize that the goal is
to compare three types of plasma (M00, M25, and M50) with
different E-charges and M-charge concentration, and all the
comparison will be made by plotting certain macroscopic properties
as a function of plasma parameter $\Gamma$.

\begin{figure}
\hspace{+0cm} \centerline{\epsfxsize=8.cm\epsffile{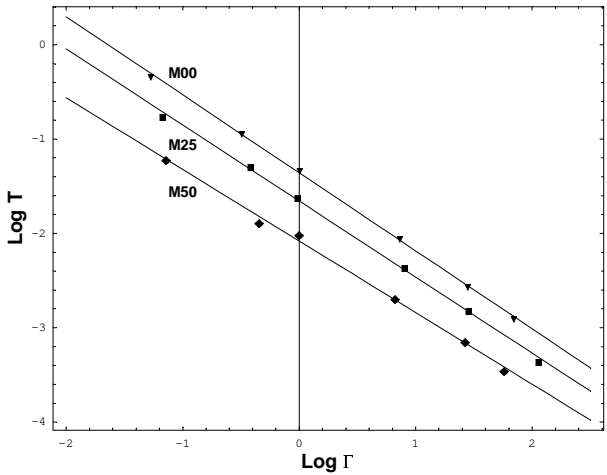}}
\hspace{+0.5cm}
 \caption{\label{fig_eos_t}
Temperature $T$ calculated at different plasma parameter $\Gamma$
in log-log plot for M00(triangle), M25(square), and M50(diamond)
plasma respectively, with the three lines from linear fitting (see
text). }
 \end{figure}

The first quantity we want to look at is the
temperature\footnote{By temperature $T$ we actually mean $k_B T$ (with the
dimension of energy in our units) throughout this paper.} dependence on
$\Gamma$ which is sort of
equation of state for plasma.\footnote{Remember in this classical statistical
system  the kinetic energy per particle is given by $E_k=3 T/2$ and
total energy per
particle is $E=(1-\Gamma)*E_k$, so the temperature dependence on
$\Gamma$ also gives all information on energy.} In
Fig.\ref{fig_eos_t} the EoS for M00, M25, and M50 are compared in
log-log plots. Data for all three show a linear relation with
similar slopes but different intercepts. By simple linear fitting
we get the following parameterized EoS for them:
\begin{eqnarray} \label{eos}
&&M00 \quad : \quad T=0.257 \, / \, \Gamma ^ {\, 0.827} \nonumber \\
&&M25 \quad : \quad T=0.191 \, / \, \Gamma ^ {\, 0.806} \nonumber\\
&&M50 \quad : \quad T=0.125 \, / \, \Gamma ^ {\, 0.759}
\end{eqnarray}
So already from the EoS we've seen considerable difference among
the three plasma. Since EoS is important for dynamical processes,
we proceed to study correlation functions and transport
coefficients in next section, expecting to see more differences.

\section{Correlation functions and transport coefficients}

Study of transport coefficients is very important in order to understand
the experimental discoveries about sQGP, such as the very low viscosity and
the diffusion of heavy quarks. In this section the transport coefficients of
our three different plasma will be calculated and compared in order to see the
influence of magnetic charges on the transport properties. To do that, we will first
measure certain correlation functions and then relate them to the corresponding transport
coefficients through the Kubo-type formulae, as is usually done in MD works.

\subsection{Velocity autocorrelation and diffusion constant}

The first correlation function  we will study is the velocity autocorrelation which is defined as:
\begin{equation} \label{vv-corr}
D(\tau) = \frac{1}{3N} < \sum_{i=1}^N \vec v_i (\tau) \cdot \vec v_i(0)  >
\end{equation}
Here $\tau$ is the correlation time, $\vec v_i$ denotes the velocity of the $i$th particles
and the sum is over all particles. The average is over thermal ensemble which is done in numerical program by average over all time points (with the number typically of order $\sim 1000$). \\

\begin{figure}
\hspace{+0.5cm} \centerline{\epsfxsize=7.5cm\epsffile{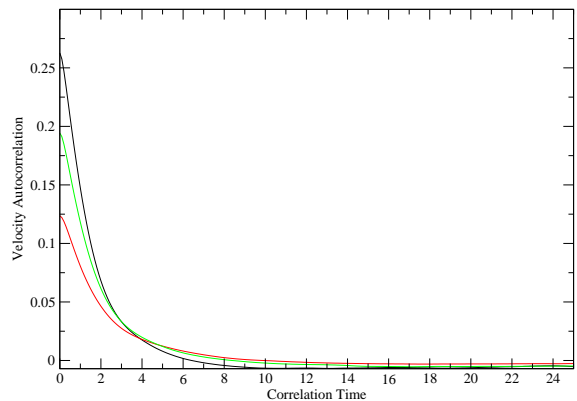}}
\hspace{+0.5cm}
 \caption{\label{fig_dift}
(color online) Velocity autocorrelation functions $D(\tau)$ for
(from top down at zero time) M00(black curve), M25(green curve)
and M50(red curve) plasma, taken at $\Gamma=1.01,0.99,1.00$
respectively. }
 \end{figure}

In Fig.\ref{fig_dift} we show typical curves for velocity autocorrelation function in M00, M25, and M50 plasma respectively. A fast damping behavior at small correlation time is observed, followed by small fluctuation from random noise at longer correlation time.

The corresponding transport coefficient, namely diffusion constant, is calculated by the following Kubo formula
\begin{equation} \label{dif-cons}
D = \int_{0}^{\infty}  D(\tau) \, d \tau
\end{equation}

\begin{figure}
\hspace{+0cm}
\centerline{\epsfxsize=7.5cm\epsffile{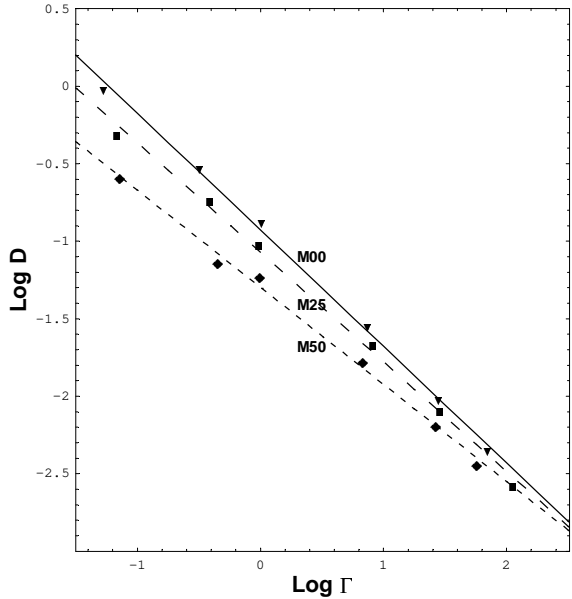}}
\hspace{+0.5cm}
 \caption{\label{fig_diffusion}
Diffusion constant $D$ calculated at different plasma parameter
$\Gamma$ in log-log plot for M00(triangle), M25(square), and
M50(diamond) plasma respectively, with the three lines from linear
fitting (see text). }
 \end{figure}

In Fig.\ref{fig_diffusion} we plot $Log D$ as a function of $Log
\Gamma$ for M00,M25 and M50 plasma. Approximate linear relation is
seen for all three, but with visible difference in slopes and
intercepts. A linear fit gives the following approximate functions
$D(\Gamma)$:
\begin{eqnarray} \label{diffusion}
&&M00 \quad : \quad D=0.396 \, / \, \Gamma ^ {\, 0.752} \nonumber \\
&&M25 \quad : \quad D=0.342 \, / \, \Gamma ^ {\, 0.707} \nonumber\\
&&M50 \quad : \quad D=0.273 \, / \, \Gamma ^ {\, 0.626}
\end{eqnarray}
At small $\Gamma<1$ there are considerable differences of the
three lines. In the physically interesting region $\Gamma \sim
1-10$ the three plasma have visible but not too much difference in
diffusion constants. The three lines will cross at about $\Gamma
\sim 10$ and after that deviation from each other again grows
quickly. The important feature common to all three types of plasma as well as to cQGP model in \cite{GSZ} is the power-law dropping of diffusion constant with increasing coupling strength. We see the diffusion constant can become few orders of magnitude smaller when one changes from weakly coupled gasous regime into strongly coupled liquid regime. This qualitative scaling in coupling is also found from AdS/CFT calculation by Casalderrey-Solana and Teaney in \cite{Casalderrey_Teaney}.

Interestingly if one combines (\ref{diffusion}) and (\ref{eos}),
the dependence of $D$ on $T$ is then obtained:
\begin{eqnarray} \label{dif_t}
&&M00 \quad : \quad D=1.36 \, T  ^ {\, 0.91} \nonumber \\
&&M25 \quad : \quad D=1.46 \, T  ^ {\, 0.88} \nonumber\\
&&M50 \quad : \quad D=1.52 \, T  ^ {\, 0.82}
\end{eqnarray}

\subsection{Stress tensor autocorrelation and shear viscosity}
\begin{figure}
\hspace{+0.5cm} \centerline{\epsfxsize=8.cm\epsffile{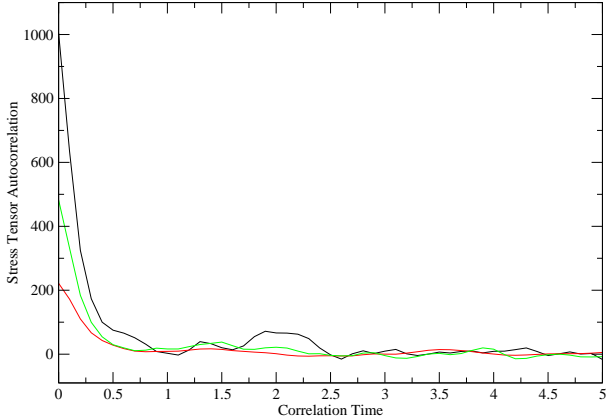}}
\hspace{+0.5cm}
 \caption{\label{fig_vct}
 (color online) Stress tensor autocorrelation functions $\eta(\tau)$ for
(from top down at zero time) M00(black curve), M25(green curve) and M50(red curve) plasma, taken at $\Gamma=1.01,0.99,1.00$ respectively.
}
 \end{figure}

It is of particular interest to study the shear viscosity of
our three plasma, as the low viscosity is one of the most important
discoveries for sQGP from RHIC experiments. For this purpose, one can
measure the autocorrelation of the off-diagonal elements of stress tensor,
namely
\begin{equation} \label{stress-corr}
\eta(\tau) = \frac{1}{3V T} <\sum_{l<k}^{1,2,3} {\cal T} _{lk}(\tau) {\cal T} _{lk}(0) >
\end{equation}
with the stress tensor off-diagonal elements
\begin{eqnarray} \label{stress-element}
{\cal T}_{l  k} &=& \sum_{i=1}^N m (\vec v_i)_l (\vec v_i)_k + \frac{1}{2} \sum_{i \ne j} (\vec r_{ij})_l (\vec F_{ij})_k \nonumber \\
&=& \sum_{i=1}^N m (\vec v_i)_l (\vec v_i)_k +  \sum_{i=1}^{N} m (\vec r_{i})_l (\vec a_{i})_k
\end{eqnarray}
In the above equations $i,j$ refer to particles while $l,k$ refer to components of three-vectors like separation, velocity and force. $\vec r_{ij}$ and $\vec F_{ij}$ are the separation and force from particle $i$ to particle $j$ respectively, while $\vec r_i,\vec v_i,\vec a_i$ are the position, velocity, acceloration of particle $i$. The equavalence of the two expressions in the second equation is discussed in great details in \cite{Allen-Tildesley}. The $V$ in the first equation is the system volume. In Fig.\ref{fig_vct} typical plots of $\eta(\tau)$ for three plasma are shown. Again the relaxation of real correlation is pretty quick and noises dominate the later time. With these correlation functions at hand the Kubo formula then leads to the following shear viscosity $\eta$:
\begin{equation} \label{viseta}
\eta = \int_0^{\infty} \eta(\tau) d\tau
\end{equation}

In general shear viscosity is a complicated property of many-body
systems, the value of which depends on many factors in a
nontrivial way. Roughly, a system with either very small $\Gamma$
(like a gas) or very large $\Gamma$ (like a solid) will have large
viscosity while a system in between (like a liquid) will have low
viscosity with a minimum usually in $\Gamma=1\sim 10$(see for
example \cite{GSZ} \cite{Ichimaru}). A qualitative explanation is
that both the particles in a gas and the phonons in a solid can
propagate very far (having a large mean-free-path) and transfer
momenta between well-separated parts, thus producing a large
viscosity, while in a liquid neither particles nor collective
modes could go far between subsequent scattering, thus making
momenta transfer very much localized and leading to a low
viscosity.\\

\begin{figure}
\hspace{+0cm} \centerline{\epsfxsize=8.cm\epsffile{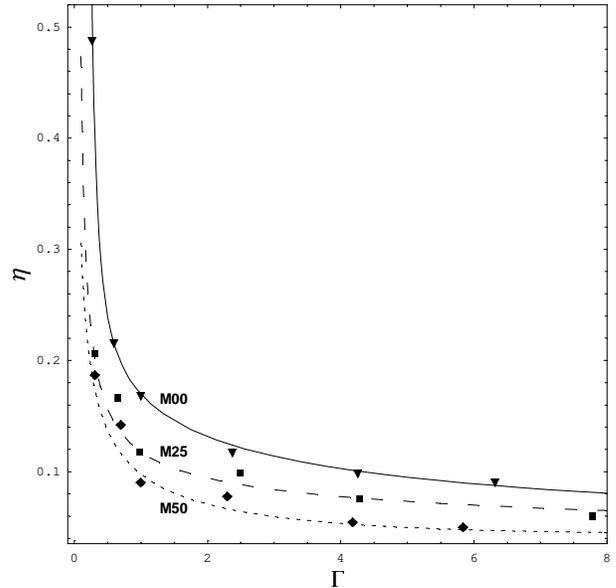}}
\hspace{+0.5cm}
 \caption{\label{fig_vis_gamma}
Shear viscosity $\eta$ calculated at different plasma parameter
$\Gamma$ for M00(circle), M25(square), and M50(diamond) plasma
respectively. }
\end{figure}

Now turning to our plasma with magnetic charges, since we have a
relatively weakly-coupled magnetic sector, one may wonder if the
magnetic particles will contribute more to large-distance momenta
transfer and hence increase the viscosity significantly. We
however argue that in the opposite, the Lorentz force induced by
the existence of magnetic charges will somehow confuse particles
and collective modes, thus helping keep the viscosity to be low.
Indeed, as shown in Fig.\ref{fig_vis_gamma}, the viscosity goes
down as increasing concentration of magnetic charges. At small
$\Gamma<1$ (in the gas phase) the three curves are getting close
to each other, but when $\Gamma$ increases into the liquid region
$\ge 1$ there is a considerable decrease of viscosity in M25 and
even more in M50 plasma. The M50 with E-charges and M-charges to
be 50\%-50\%, has the values of viscosity about half of the pure
electric M00 plasma at the same $\Gamma$. So, we conclude that the
existence of magnetic charges may help us to understand the
extremely low viscosity of sQGP. A rough parametrization of the
data gives the following viscosity dependence on $\Gamma$ in the
plotted region:
\begin{eqnarray} \label{viscosity-gamma}
&&M00 \quad : \quad \eta=0.002 \, / \, \Gamma ^ {\, 3.64} + 0.168 \, / \, \Gamma ^ {\, 0.353}\nonumber \\
&&M25 \quad : \quad \eta=0.013 \, / \, \Gamma ^ {\, 1.36} + 0.105 \, / \, \Gamma ^ {\, 0.237}\nonumber\\
&&M50 \quad : \quad \eta=0.096 \, / \, \Gamma ^ {\, 0.500}+ 0.001
\, \cdot \, \Gamma ^ {\, 1.12}
\end{eqnarray}
In all of them the first term is most dominant at very small
$\Gamma$ while the second term becomes important at relatively
large $\Gamma$. We noticed that for M50 there is already positive
power term of $\Gamma$, which is in accord with the expected
qualitative feature. Similar terms will appear in two other plasma
when we will be able to include in the fitting more points from
large $\Gamma$.

\subsection{Electric current autocorrelation and conductivity}

The last transport property we study in this paper is the electric current autocorrelation and the electric conductivity. This analysis is only done for pure electric M00 plasma since the comparison among M00, M25 and M50 (which already have different E-charge concentrations) doesn't make much sense. The electric current autocorrelation is given by
\begin{equation} \label{jj-corr}
\sigma(\tau) = \frac{1}{3VT} < (\sum_{i=1}^N e_i \vec v_i (\tau))
\cdot (\sum_{i=1}^N e_i \vec v_i(0) ) >
\end{equation}
with $e_i$ the electric charge of the $i$th particle. And the electric conductivity is obtained from Kubo formula as
\begin{equation} \label{conductivity}
\sigma = \int_{0}^{\infty} \sigma(\tau) d \tau
\end{equation}

\begin{figure}
\hspace{+0cm} \centerline{\epsfxsize=7.cm\epsffile{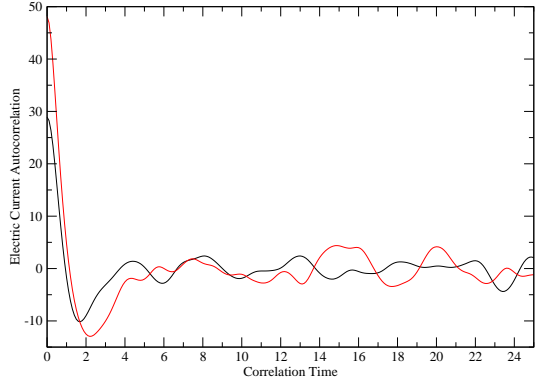}}
\hspace{+0.5cm}
 \caption{\label{fig_jjct}
(color online)  Electric current autocorrelation functions
$\sigma(\tau)$ for pure electric M00 plasma taken at (from top
down at zero time) $\Gamma=6.33$(red curve) and
$\Gamma=14.53$(black curve). }
 \end{figure}

In Fig.\ref{fig_jjct} we show the typical $\sigma(\tau)$ as a
function of $\tau$ for two values of $\Gamma$. For this
correlation function we do notice that even for $\Gamma$ not
large, the late time correlation is not purely noise but still has
small oscillation. This is not unreasonable since related
collective modes like plasmon may develop even for a gas. After
integration it turns out in the region $\Gamma \sim 0.3-15$ the
conductivity is scattered between $\sigma=0.101-0.141$ without
clear tendency, which may indicate the electric current
dissipation is not sensitive to $\Gamma$ in this region. It is
very interesting to see what will happen to the color-electric
conductivity (giving information about color charge transport) in
a Non-Abelian plasma.\\

\subsection{Mapping between MD systems and sQGP} \label{mapping}

With the MD-obtained empirical formulae for diffusion and viscosity,
it is of great interest to see what they predict for the parameter region
corresponding to the sQGP experimentally created at RHIC. To do this
mapping, one has to identify the corresponding physical values of
basic units (namely mass, length and time) in the destination system and
then combine dimensionless numbers and relations from MD with proper
dimensions. Also the plasma parameter $\Gamma$ should be determined for the
destination system such that we pick up the MD-predicted values of
interesting quantities (say, diffusion constant and shear viscosity) at
exactly the same $\Gamma$ value.

Following similar estimates as in \cite{GSZ}, we summarize below the relevant
quantities of sQGP around $1.5T_c$:\\
1. Quasiparticle (quarks and gluons) mass can be estimated as $m \approx 3.0 T$;\\
2. The typical length scale is simply estimated from quantum
localization to be
$r_0 \approx 1 / m  \approx 1/(3.0 T)$;\\
3. The electric coupling strength, after averaging over different EQPs(quarks and gluons)
with their respective Casimir, is roughly $<\alpha_s C> \approx 1$; \\
4. The particle density is roughly given, under the light that lattice results
have shown the sQGP pressure and entropy to reach about 0.8 of Stefan-Boltzmann limit, by $n \approx 0.8(0.122\times 2\times 8+ 0.091 \times 2\times 2\times N_c\times N_f)T^3 \approx 4.2 T^3 $;\\
5. This density estimation leads to the Wigner-Seitz radius $a_{WS}=(\frac{3}{4\pi n})^{1/3}\approx 1/(2.6 T) \approx 1.1 r_0$;\\
6. We then get the time scale as the inverse of plasmon frequency $\tau_p=1/\omega_p = (\frac{m}{4\pi n \alpha_s C})^{1/2}\approx 1/(4.2T)$;\footnote{This $\tau_p$ has subtle difference in time scale used in our MD, namely the MD time unit $\tau$ is related to inverse of plasmon frequency by $\tau=\tau_p\times (4\pi n \lambda^3)^{1/2} \approx 1.46 \tau_p$ which should be taken into account for mapping.}\\
7. The entropy density is estimated from Stefan-Boltzmann limit as $s \approx 0.8 \times \frac{4\pi^2}{90} [2\times 8 + (7/8)\times 2\times 2 \times N_c\times N_f]T^3 \approx 16T^3 $.

Now let's discuss the value of $\Gamma$. As already mentioned, the $\Gamma$ given in our MD is the actual ratio of potential to kinetc energy, which is measured during the simulation. The usually quoted one, defined as $\tilde{\Gamma}=\frac{e^2}{a_{WS} (k_B T)}$, could be considered as a pre-determined 'superficial Gamma'. Unfortunately it is not clear how to estimate the actual Gamma $\Gamma$ of sQGP while the superficial Gamma $\tilde{\Gamma}$ is obtainable for sQGP, which is $\tilde{\Gamma} \approx 2.6 <\alpha_s C> \approx 2.6$. So we should try to figure out the superficial Gamma in our MD and map the results accordingly.  The two are different though, they are monotonously related to each other, namely when one is large(small) so is the other. Since our MD has been done with $n \lambda^3 = 0.17$, the $(a_{WS})_{MD}\approx 1.12 \lambda$ which means in our MD $\tilde{\Gamma}= 0.89 / T$, which after combination with (\ref{eos}) will give us the convertion formula between the two Gamma's. Similar convertion relation could also be obtained for cQGP in \cite{GSZ} from their Fig.8 though in their case they use superficial Gamma as basic parameter and measure potential energy from simulation.

\begin{figure}
\hspace{+0cm} \centerline{\epsfxsize=7.cm\epsffile{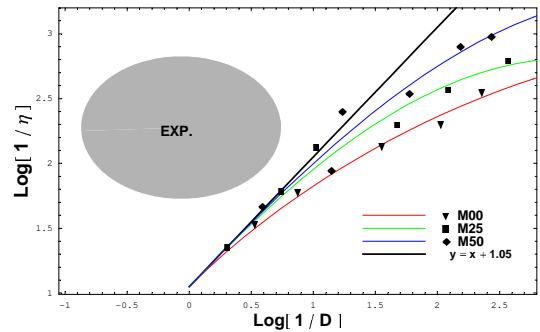}}
\hspace{+0.5cm}
 \caption{\label{fig_vis_d_MD}
(color online) Plots of $Log[1/\eta]$ v.s. $Log[1/D]$ for three different plasmas. The shaded region is
mapped back from experimental values, see text. }
 \end{figure}

With all the above ingredients we are at place to do the mapping
for interesting transport coefficients $D$ and $\eta$ between our
MD systems and the sQGP. The mapping is a two-way business: one
may map the experimentally suggested values back into
corresponding MD numbers, as is done and shown in
Fig.\ref{fig_vis_d_MD}; or one can use the MD-obtained relations
to predict the corresponding relations of sQGP after conversion of
units, as is shown in Fig.\ref{fig_vis_d_mapping} and discussed in
the summary part. Here let's focus on MD systems in
Fig.\ref{fig_vis_d_MD} where a $Log[1/ \eta]$ v.s. $Log[1/ D]$ is
plotted: data points for all three plasmas fall on a universal
unit-slope straight line on the left lower part, indicating a
small $\Gamma$ gas limit with diffusion and viscosity both
proportional to mean free path; all three curves soon deviate from
gas limit at larger $\Gamma$ (strong coupling) and become flat in
the liquid region; the shaded oval is obtained by mapping back the
following experimental values: $\eta/s \approx 0.1-0.3$, $2\pi T D
\approx 1-5$, which is clearly not close to gas region but near
the liquid region, especially the one of the M50 curve. More about
comparison will be given in the summary part at the end of the
paper.\\

\section{Collective excitations at very strongly coupled regime}

In this section we will report interesting results for collective
excitations found at very strongly coupled regime ($\Gamma$
greater than a few tens) of the pure electric (M00) plasma. The
signals of these excitations are extraordinarily clear when
$\Gamma$ goes to $\sim 100$ or larger. We came to notice these
very good modes not in a straight forward way. Instead, these
modes have revealed themselves dramatically in
 some unusual structures of the dynamical correlation functions and their
 fourier spectra, which we measured first. Only after thinking
 about possible source of these structures we turned to
 systematic and direct measurements for certain collective modes, which are
 found to coincide with correlation functions' structures in a distinct manner.
As mentioned before, in this regime our plasma is like a
"self-holding drop" which has very different collective motions
from a plasma in periodic boxes: the latter has the familiar
phonon modes while the former is really like a raindrop, having
vibration modes like monopole modes, dipole modes, quadruple
modes, etc corresponding to different components of density
distribution's spherical harmonics. We will discuss these modes
respectively in more details in the following.\\

\subsection{Monopole modes}

\begin{figure}
\hspace{+0.cm}
\centerline{\epsfxsize=8.cm\epsffile{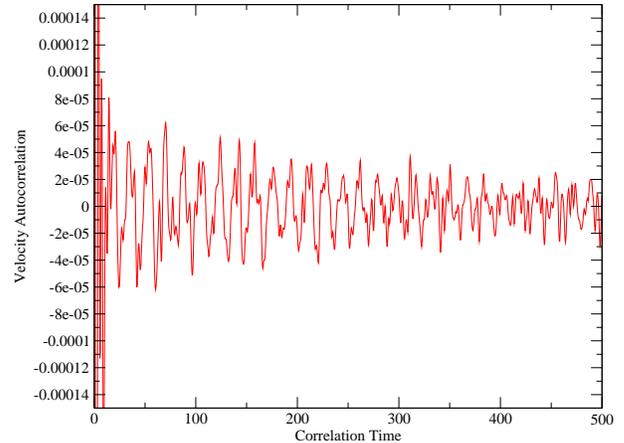}}
\hspace{+0.5cm}
 \caption{\label{fig_vvt}
(color online) Velocity autocorrelation function taken at $\Gamma=116.91$.
}
 \end{figure}

\begin{figure}
\hspace{+0.cm}
\centerline{\epsfxsize=8.cm\epsffile{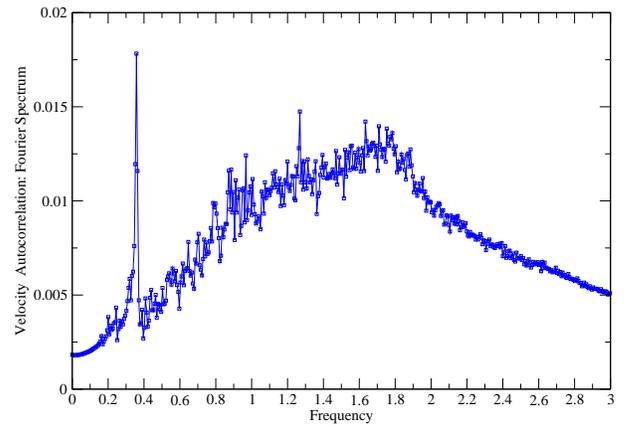}}
\hspace{+0.5cm}
 \caption{\label{fig_vvw}
(color online)  Fourier transformed spectrum of velocity autocorrelation function taken at $\Gamma=116.91$.
}
 \end{figure}

Let's start with the velocity autocorrelation function
(\ref{vv-corr}) which is supposed to almost vanish (except a
little random noise) at large correlation time and give convergent
integral to yield diffusion constant, as seen in previous section.
However when measured in the very strongly coupled regime, this
correlation function is found to have robust oscillating behavior
even for very large time which of course couldn't be considered as
noise, see Fig.\ref{fig_vvt}. By looking at the fourier spectrum
of it, one immediately sees a large and narrow peak at
$\omega^D_1=0.35$ very clearly on top of a very broad shoulder
structure, as shown in Fig.\ref{fig_vvw}. These behaviors are true
for $\Gamma$ down to about $50$.\\ 

Now the question is why there will be such peaks in velocity
autocorrelation. The answer lies in the monopole modes, which can
be directly measured through simply the time dependence of
average particle radial position, namely:
\begin{equation} \label{monopole_mode}
 R (t)  = \frac{1}{N} \sum_{i=1}^N |\vec r_i (t)|
\end{equation}

\begin{figure}
\hspace{+0.cm}
\centerline{\epsfxsize=8.cm\epsffile{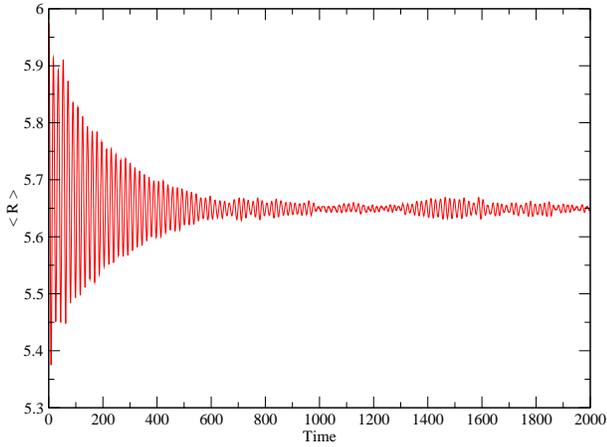}}
\hspace{+0.5cm}
 \caption{\label{fig_rt}
(color online) Average monopole moment $R(t)$ (see text) as a function of time, taken at $\Gamma=116.91$.
}
 \end{figure}

\begin{figure}
\hspace{+0.cm}
\centerline{\epsfxsize=8.cm\epsffile{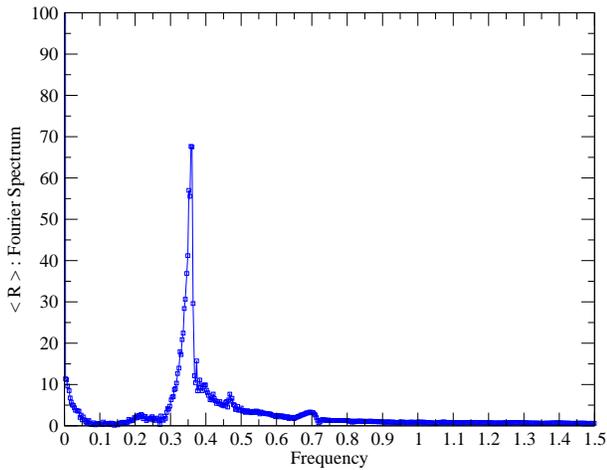}}
\hspace{+0.5cm}
 \caption{\label{fig_rw}
(color online) Fourier transformed spectrum of average monopole moment $R(t)$(see text), taken at $\Gamma=116.91$.
}
 \end{figure}

In Fig.\ref{fig_rt} one sees really nice oscillations lasting for
near hundred revolutions. In a small movie showing the positions
of particles at several subsequent time points the "drop" in
monopole mode looks like a beating heat. This oscillation
amplitude decreases slowly indicating a nonzero but small width of
this monopole mode. Again the fourier spectrum gives important
information such as the characteristic frequency and width of
collective mode. In Fig.\ref{fig_rw} we can see one major narrow
peak at $\omega^M_1=0.35$ together with a few roughly visible but
much smaller lumps at $\omega=0.22,0.46,0.70$. $\omega=0.70$
structure may be a secondary harmonics of the major peak, and
seemingly the $0.22$ and $0.46$ may also be in the same series of
harmonics with different ranks.

The important finding we want to point out is the coincidence of
$\omega^M_1$ here with the peak $\omega^D_1$ from velocity
autocorrelation function. This result tells us the monopole mode,
in a form of radial vibration, has nontrivial influence on
velocity autocorrelation and consequently on particle diffusion.\\

\subsection{Quadruple modes}

\begin{figure}
\hspace{+0.cm}
\centerline{\epsfxsize=7.5cm\epsffile{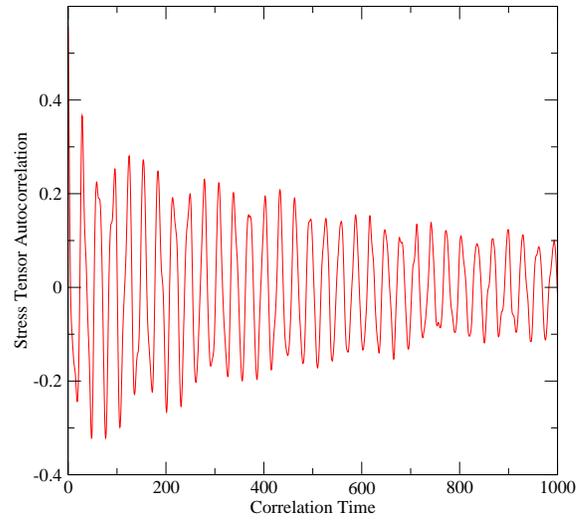}}
\hspace{+0.5cm}
 \caption{\label{fig_vist}
 (color online) Stress tensor autocorrelation function taken at $\Gamma=116.91$.
}
 \end{figure}

\begin{figure}
\hspace{+0.cm}
\centerline{\epsfxsize=7.5cm\epsffile{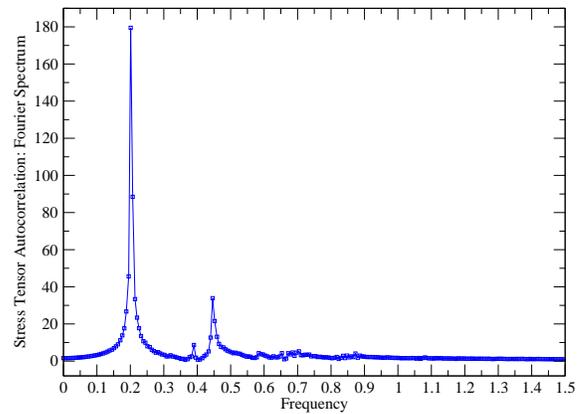}}
\hspace{+0.5cm}
 \caption{\label{fig_visw}
(color online) Fourier transformed spectrum of stress tensor autocorrelation function taken at $\Gamma=116.91$.
}
 \end{figure}

The study of stress tensor autocorrelation in the strongly coupled
plasma gives us even more interesting correspondence between
correlation functions and collective modes. In Fig.\ref{fig_vist}
we plot the stress tensor autocorrelation (\ref{stress-corr}) as a
function of time, and in Fig.\ref{fig_visw} its fourier spectrum,
in which three clear and narrow peaks can be seen at
$\omega^{\eta}_1=0.20$, $\omega^{\eta}_2=0.40$, and
$\omega^{\eta}_3=0.45$. The $0.40$ peak, which is the smallest
one, may be a secondary harmonics of the remarkable $0.20$ peak.
At $\Gamma$ as small as about 25, the $0.20$ peak is still alive
in this correlation function. Because of the existence of these,
the correlation function has significant oscillations with large
amplitude even for very large correlation time.\\ \\

 \begin{figure}
\hspace{+0.cm}
\centerline{\epsfxsize=8.cm\epsffile{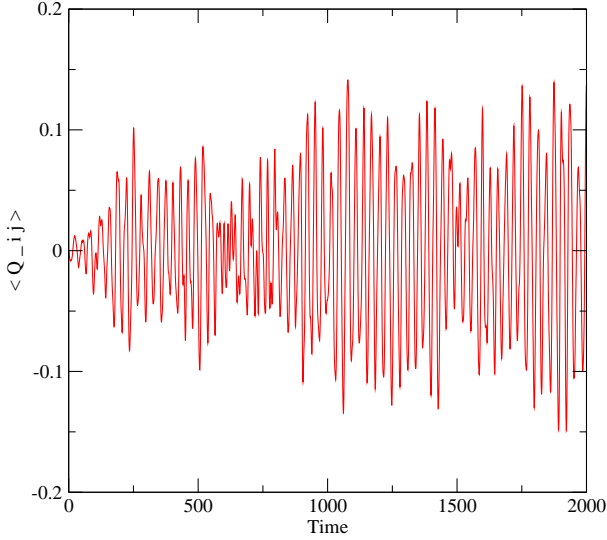}}
\hspace{+0.5cm} \vspace{0.1in}
 \caption{\label{fig_qijt}
(color online)  Average off-diagonal quadruple moment $Q_{23}(t)$ (see text)
as a function of time, taken at $\Gamma=116.91$.
}
 \end{figure}

\begin{figure}
\hspace{+0.cm}
\centerline{\epsfxsize=8.cm\epsffile{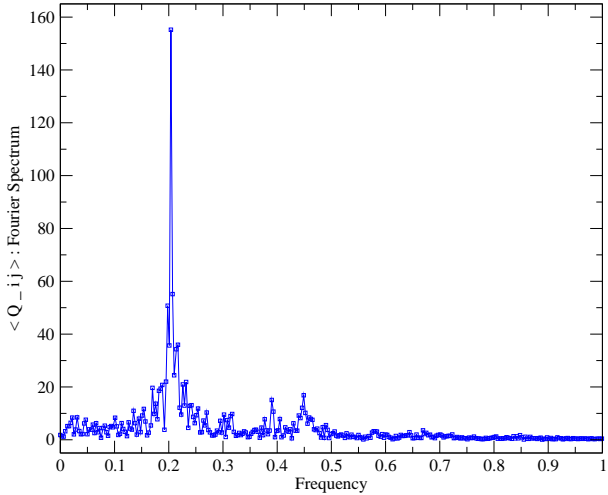}}
\hspace{+0.cm} \vspace{0.3in}
 \caption{\label{fig_qijw}
(color online) Fourier transformed spectrum of average
off-diagonal quadruple moment $Q_{23}(t)$ (see text), taken at
$\Gamma=116.91$. }
 \end{figure}

To find the source of these, we directly measured the off-diagonal
quadruple modes by the following probe:
\begin{equation} \label{off_diag_quadruple}
Q_{lk} (t) =  \frac{1}{N} \sum_{i=1}^{N} (\vec r_i (t))_l (\vec
r_i(t))_k \quad  , \,  l,k=1,2,3, \quad l \ne k
\end{equation}

We have three independent of them, say $Q_{12},Q_{23},Q_{31}$. In
Fig.\ref{fig_qijt} we show one of them as a function of time with
similar results for the other two. From the figure we can see that
at the very beginning there is almost no quadruple mode but its
amplitude grows significantly in a time interval $0-200$ during
which the monopole mode decays down (see Fig.\ref{fig_rt}). Then
after that they persist for long time. This indicates that these
off-diagonal quadruple modes are very robust and somehow "cheap"
to excite and the energy initially in the monopole modes is
preferably transferred into the quadruple modes.

Now when we plot the fourier spectrum of the off-diagonal
quadruple modes in Fig.\ref{fig_qijw}, amazingly three very clear
peaks appear at $\omega^Q_1=0.20$, $\omega^Q_2=0.40$, and
$\omega^Q_3=0.45$, which are exactly the same frequencies found in
the stress tensor autocorrelation. The relative amplitudes among
the three peaks are also similar in two cases. This is a profound
correspondence which means the stress tensor correlation and the
related transport property, namely viscosity, are especially
dominated by the off-diagonal quadruple modes of the system. This
type of connection may be universal and one may find certain
collective excitations for each dynamical correlation function.

Before closing this subsection, let's also mention the diagonal
quadruple modes which we also studied. This part of quadruples
could be probed by the following quantity:
\begin{eqnarray} \label{diagonal_quadruple}
&& Q_{ll} (t) = \frac{1}{N} \sum_{i=1}^{N}  {\big [ } 3 ((\vec
r_i(t))_l)^2 - |\vec r_i(t)|^2 {\big ]} \quad , \, l=1,2,3 \nonumber \\
&&
\end{eqnarray}

It has similar behavior as the off-diagonal modes (to save space
we skip to show the plots) , with peaks in spectrum at
$\omega^Q_4=0.12$ and $\omega^Q_5=0.25$ which presumably are
different ranks in the same harmonic series. These peaks however
are not seen in the stress tensor autocorrelation, which is
understandable since the stress tensor autocorrelation we study is
actually the correlation of stress tensor's off-diagonal parts
which is related to shear viscosity. We think these peaks of
diagonal quadruple modes must be seen in the diagonal parts of
stress tensor correlation which is related to the bulk viscosity.\\ \\

\subsection{Plasmon modes}

It is clear that our "drop" won't have dipole (or any
odd-multiple) excitation according to symmetric setting. But there
can be another type of dipole excitations, namely the electric
dipole modes, or in a more common notion the plasmon modes. These
can be probed by
\begin{equation} \label{e_dipole}
(e\vec R)_l (t) = \frac{1}{N} \sum_{i=1}^{N} e_i (\vec r_i (t))_l
\quad , \, l=1,2,3
\end{equation}
\\

\begin{figure}
\hspace{+0.cm}
\centerline{\epsfxsize=8.cm\epsffile{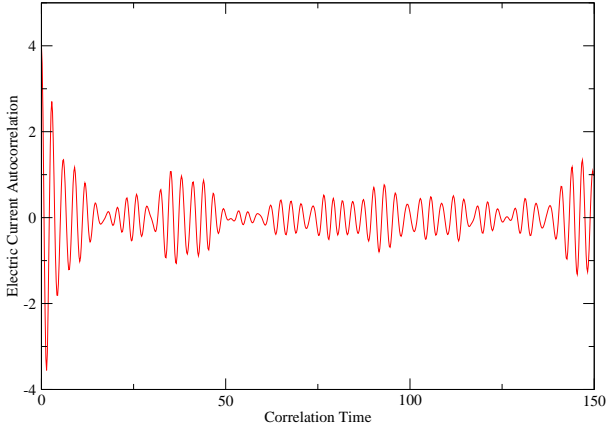}}
\hspace{+0.5cm}
 \caption{\label{fig_jet}
(color online) Electric current autocorrelation function taken at
$\Gamma=116.91$. }
 \end{figure}

\begin{figure}
\hspace{+0.cm}
\centerline{\epsfxsize=8.cm\epsffile{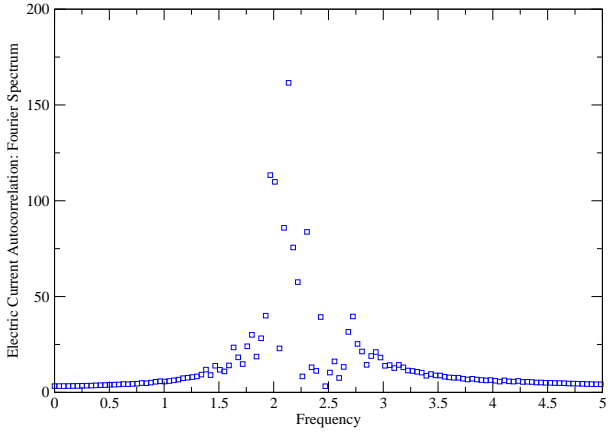}}
\hspace{+0.5cm}
 \caption{\label{fig_jew}
(color online) Fourier transformed spectrum of electric current
autocorrelation function taken at $\Gamma=116.91$. }
 \end{figure}

 \begin{figure}
\hspace{+0.cm}
\centerline{\epsfxsize=8.cm\epsffile{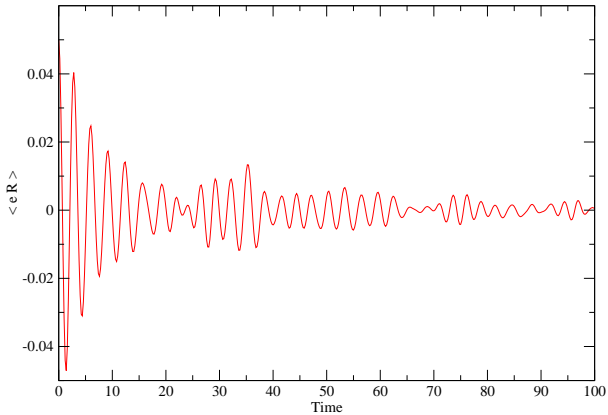}}
\hspace{+0.5cm}
 \caption{\label{fig_ert}
(color online)  Average electric dipole moment $eR(t)$ (see text)
as a function of time, taken at $\Gamma=116.91$. }
 \end{figure}

\begin{figure}
\hspace{+0.cm}
\centerline{\epsfxsize=8.cm\epsffile{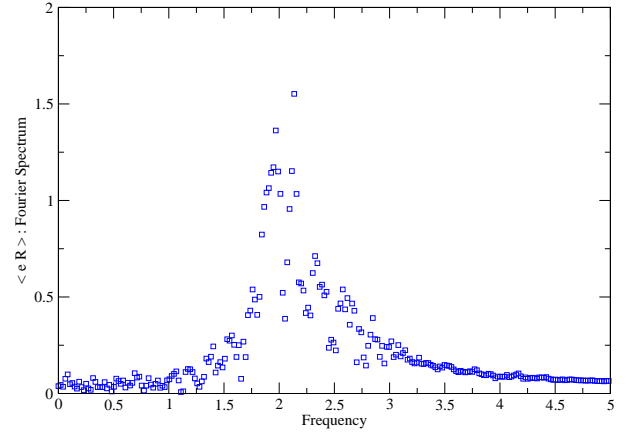}}
\hspace{+0.5cm}
 \caption{\label{fig_erw}
(color online) Fourier transformed spectrum of average electric
dipole moment $eR(t)$ (see text), taken at $\Gamma=116.91$. }
 \end{figure}

The corresponding correlation function should be the electric
current autocorrelation (\ref{jj-corr}). To clearly reveal these
modes we shift our positive charges and negative charges with
small displacement in opposite directions at the initial time
which introduces zero dipole but nonzero electric dipole. We then
plot the electric current autocorrelation (Fig.\ref{fig_jet}) and
its fourier spectrum (Fig.\ref{fig_jew}), and the direct electric
dipole (Fig.\ref{fig_ert}) and its fourier spectrum
(Fig.\ref{fig_erw}) as well.

Again one see similar behavior in both and find similar peak
structure at $\omega^J=\omega^E=2.0$ in both spectra. These peaks
are large but broad and have some fluctuation, as compared with
previous peaks, which is understandable as the plasmon modes
usually have bigger width (larger dissipation) than sound modes.
These modes seem to be present even if the system has only $\Gamma
\sim 10$. If one calculate the plasmon frequency using the simple
formula $\omega_p=\sqrt{\frac{4\pi n e^2}{m}}$ for our drop (with
$n=1/a^3 \, , \, a=1.18\lambda$) we then get $\omega_p=2.7$. This
is not far from the observed $2.0$ and the discrepancy must be
there because the size of our system is only about 10 times the
microscopic scale and the positive and negative charges in the
middle are not entirely screening each other as assumed when
deriving the formula.

\subsection{Size scaling of the collective modes}

\begin{figure}
\hspace{+0.cm} \centerline{\epsfxsize=8.cm\epsffile{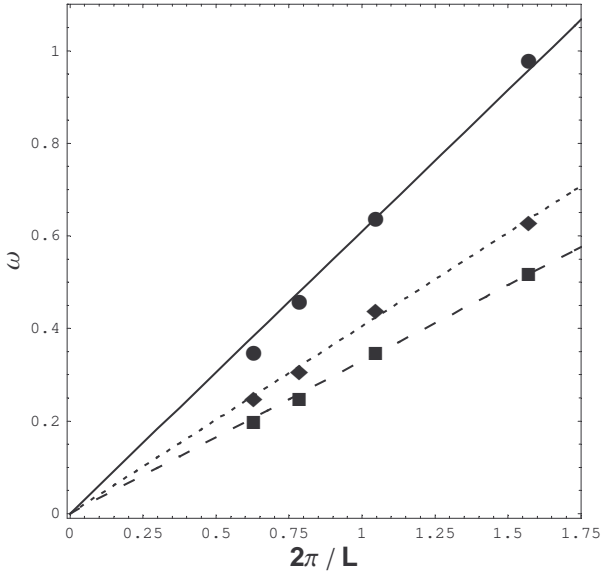}}
\hspace{+0.5cm}
 \caption{\label{fig_dispersion}
Peak frequency as a function of $2\pi / L$ (with $L$ the system size) for
monopole modes(circle), diagonal quadruple modes(diamond), and off-diagonal quadruple modes(square) respectively, with the three lines from linear fitting (see text).
}
 \end{figure}

Since the monopole and quadruple modes should be sound modes, it
is interesting to see how their frequencies scale with the system
size. For a large enough system one expects the sound modes
dispersion to be $\omega=c_s k - \frac{i}{2} \Gamma_s k^2$, namely
the mode frequency itself scales linearly in $k$ while the width
scales quadratically. What we did is to change system size to be
$10\times 10 \times 10$, $8\times 8 \times 8$, $6\times 6 \times
6$, and $4\times 4 \times 4$, and then look at the change of peaks
in those monopole and quadruple modes (picking the major peaks
$\omega^M_1$, $\omega^Q_1$ and $\omega^Q_5$). As demonstrated in
Fig.\ref{fig_dispersion} where these frequencies are shown as
function of $2\pi / L$ with $L$ the system size, the linear
scaling is very well observed. We obtain the following linear
fitting for the three modes:
\begin{eqnarray} \label{scaling}
Monopole\, &&: \,  \omega^M_1 = 0.610 \, \cdot \, k \nonumber \\
Quadruple \, D \, && : \,  \omega^Q_5 = 0.404 \, \cdot \, k  \nonumber \\
Quadruple \, N-D \, && : \,  \omega^Q_1 = 0.329 \, \cdot \, k
\end{eqnarray}
The two quadruple modes have similar slope (with the diagonal one
a little larger) which means they have close propagation velocity,
while the monopole modes have a slope or propagation velocity
larger by a factor about $1.5-2$. This is reasonable, just like in
usual solid the longitudinal sound waves have larger velocity than
the transverse ones. The lines show remarkable consistency with
the fact that sound modes with infinitely large wavelength should
have zero frequency. For the width however we didn't unambiguously
see a regular dependence on $L$, which indicates our systems are
not macroscopic enough since the width is more sensitive than
frequency itself to the dissipation effect related to system size.

The last result to mention is about the dependence of plasmon
modes on system size. Actually we found for all four sizes, the
plasmon frequencies roughly stay around $\omega^E=2.0$ without
changing, only with the peak structure getting worse. Again this
is expected since the usual plasmon dispersion displays a plateau
at $\omega_p=\sqrt{\frac{4\pi n e^2}{m}}$ for $k$ not large.

\section{Summary and discussion}

In summary, we propose to view the finite-$T$ QCD as a competition
between electrically charged quasiparticles (EQPs) and the
magnetically charged  (MQPs). The high-$T$/high density
 limit is known to be perturbative QGP, which is
electric-dominated. This implies that EQPs are more numerous, with
density $\sim N_c^2 T^3$, while the density of MPQs is $\sim N_c^2
T^3/log^3(T/\Lambda_{QCD})$. In this case the electric coupling is
$weaker$ than the magnetic $e<g$. We think that at some
intermediate $T\sim 300 \, MeV$ both sectors' couplings and
densities are similar, and below it $T<300 \, MeV$ the roles are
reversed, with dominant MQPs and electric coupling being
$stronger$ than magnetic $e>g$. One of the important consequences
of this picture is ``postconfinement'' phenomena and
(electrically) strongly coupled QGP right above the deconfinement
phase transition $T= (200-300)\, MeV$.

Using these ideas as a motivation, we start a program of studies
aimed at understanding the many-body aspect of matter composed of
both of electrically and magnetically charged quasiparticles.
(Two-body charge-monopole and charge-dyon situations have been
covered in the literature \cite{Goldhaber,Milton}.)

Beginning with a 3 body problem: a static E-dipole plus a
dynamical monopole, we have found evidences both classically and
quantum mechanically that a monopole can be bound to an electric
dipole, which will later be more thoroughly treated in a dilute
monopole gas scenario and may eventually lead to an explanation of
large entropy associated with static quark-anti-quark just above
$T_c$ as indicated by lattice data.

We have then used molecular dynamics to do the first systematic
study of a plasma with both electric and magnetic charges. Two
regimes have been separately studied:\\
1. In the weak and medium coupling regime ($\Gamma<25$), which is
most relevant to sQGP problem, the equation of state, the
correlation functions and the transport coefficients have been
evaluated and compared among plasma with different magnetic
contents. Most interestingly we found by increasing the
concentration of magnetic charges to about 50\% we get a factor
$2$ down for viscosity, which is particularly important in view
of explaining  surprisingly low
viscosity of sQGP as observed at RHIC.\\
2. In the very strongly coupled regime ($\Gamma>25$), very
interesting correspondence between correlation functions and
collective excitations have been revealed. The monopole modes are
found to induce long time oscillation in velocity autocorrelation,
while the off-diagonal quadruple modes are shown to have profound
dominance on the nontrivial oscillating behavior of stress tensor
(off-diagonal) autocorrelation. In a similar manner the plasmon
modes are connected to electric current autocorrelation. So with
such relation, studying one side also gives us information about
the other side. The sound modes nature of such collective
excitations has been demonstrated by studying the scaling property
of their peak frequencies with system size.

\begin{figure}
\hspace{+0cm} \centerline{\epsfxsize=8.cm\epsffile{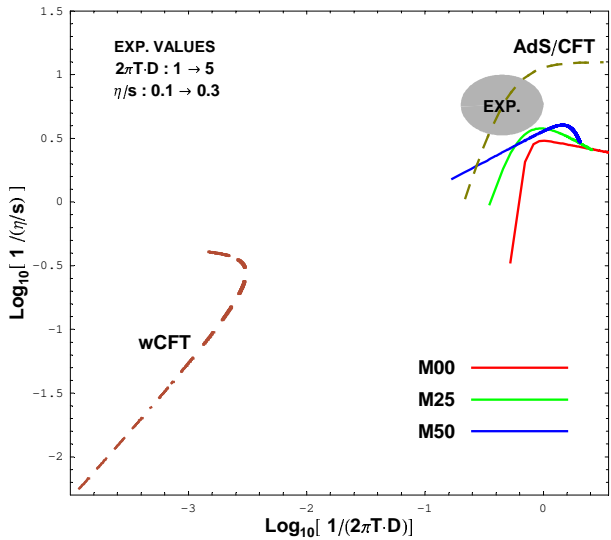}}
\hspace{+0.5cm}
 \caption{\label{fig_vis_d_mapping}
(color online) Plots of $Log[1/(\eta /s)]$ v.s. $Log[1/(2\pi T D)]$ including
results from our MD simulations, the Ads/CFT calculations, the weakly coupled
CFT calculations, as compared with experimental values, see text. }
 \end{figure}

Finally, we would like to compare our results with those
obtained using the AdS/CFT correspondence and also with empirical data
about sQGP from RHIC experiments. Those are summarized in
Fig.\ref{fig_vis_d_mapping}, as a log-log plot of properly normalized
dimensionless (heavy quark) diffusion constant and viscosity.

The dashed curve in the left lower corner is for $\cal N$=4 SUSY
YM theory in weak coupling, where viscosity is from
\cite{Huot:2006ys} and diffusion constant
from\cite{Chesler:2006gr}. The curve has a slope of one on this
plot, as in weak coupling both quantities are proportional to the
same mean free path. As one can easily see, weak coupling results
are quite far from empirical data from RHIC, shown by a gray oval
in the right upper corner. Viscosity estimates follow from
deviations of the elliptic flow at large $p_t$ from hydro
predictions \cite{Teaney_hydro}, and diffusion constants are
estimated from $R_{AA}$ and elliptic flow of charm
\cite{Teaney_diffusion}.

The curve for strong-coupling AdS/CFT results (viscosity according
to \cite{visc} with $O(\lambda^{-3/2})$ correction, diffusion
constant from \cite{Casalderrey_Teaney}), shown by upper dashed
line, is on the other hand going right through the empirical
region. At infinite coupling this curve reaches $s/\eta=4\pi$
which is conjectured to be the ever possible upper bound.

Our results -- three solid lines on the right -- correspond to our
calculations with different EQPs/MQPs ratio. They are close to the
empirical region, especially the version with the equal mixture of
EQPs and MQPs.

Let us end with a warning, that the empirical data, the mapping
from classical system to sQGP and the relation between QCD and the
$\cal{N}$ =4 SUSY YM used in AdS/CFT will only become
 quantitative with time: this figure is just the first attempt
to get together all three ingredients of the broad picture.

\vskip .25cm
{\bf Acknowledgments.}
\vskip .2cm

This work was supported in parts by the US-DOE grant DE-FG-88ER40388.
ES thanks A.Vainshtein, E. M.Ilgenfritz and C.Korthals-Altes for
helpful discussions.

\end{document}